\documentclass{jfm}

\usepackage{graphicx} 
\usepackage{natbib} 
\usepackage{subfigure}
\usepackage{amsmath,amssymb} 
\usepackage{color}
\usepackage{enumerate}
\usepackage{url}

\def\Ub{{U_b}} \def\Vb{{V_b}}

\newcommand{\Fr}{\ensuremath{\mathrm{Fr}}}

\title[Bulldozing of granular material]{
  Bulldozing of granular material}

\author[A. Sauret, N. J. Balmforth, C. P. Caulfield \& J. N.
McElwaine]%
{A. Sauret$^{1,\thanks{Email address for correspondence:
    asauret@princeton.edu}}$ ,\ns N. J. Balmforth$^{2}$, C. P.
  Caulfield$^{3,4}$ \\ \& J. N. McElwaine$^{5,6}$}

\affiliation{$^1$ Department of Mechanical and Aerospace Engineering, Princeton University, Princeton, New Jersey 08544, USA,\\[\affilskip]
  $^2$ Department of Mathematics, University of British Columbia, Vancouver, BC V6T 1Z2,  Canada\\[\affilskip]
  $^3$ BP Institute, University of Cambridge, Madingley Road, Cambridge CB3 0EZ, UK,\\[\affilskip]
  $^4$ Department of Applied Mathematics and Theoretical Physics,
  University of Cambridge,
  Centre for Mathematical Sciences, Wilberforce Road, Cambridge CB3 0WA, UK,\\[\affilskip]
  $^5$Department of Earth Sciences, University of Durham, Durham, DH1
  3LE, UK, \\[\affilskip]$^6$ WSL Institute for Snow and Avalanche
  Research SLF, Fl\"{u}elastrasse 11, Davos, Switzerland }

\date{\today}
\begin{document}

\maketitle

\begin{abstract}
We investigate the bulldozing motion of a granular sandpile  driven forwards by a vertical plate. The problem is set up in the  laboratory by emplacing the pile on a table rotating underneath a
  stationary plate; the continual circulation of the bulldozed
  material allows the dynamics to be explored over relatively long
  times, and the variation of the velocity with radius permits one to
  explore the dependence on bulldozing speed within a single
  experiment. We measure the time-dependent surface shape of the dune for a
  range of rotation rates, initial volumes and radial
  positions, for four granular materials, ranging from glass spheres
  to irregularly shaped sand. The evolution of the dune can be separated
  into two phases: a rapid initial adjustment to a state of
  quasi-steady avalanching perpendicular to the blade, followed by a
  much slower phase of lateral spreading and radial migration.  The
  quasi-steady avalanching sets up a well-defined perpendicular profile
  with a nearly constant slope. This profile can be scaled
  by the depth against the bulldozer to collapse data from
  different times, radial positions and experiments onto common
  `master curves' that are characteristic of the granular material and
  depend on the local Froude number. The lateral profile of the dune
  along the face of the bulldozer varies more gradually with radial
  position, and evolves by slow lateral spreading.
  The spreading is asymmetrical, with the inward
  progress of the dune eventually arrested and its bulk migrating to
  larger radii.  A one-dimensional depth-averaged model 
  recovers the nearly linear perpendicular
  profile of the dune, but does not capture the finer nonlinear details
  of the master curves. A two-dimensional version of the model
  leads to an advection-diffusion equation that reproduces
  the lateral spreading and radial migration.  Simulations using the
  discrete element method reproduce in more quantitative detail many
  of the experimental findings and furnish further insight into the flow
  dynamics.
\end{abstract}


\section{Introduction}

The dynamics of dense granular media plays a key role in a variety of
engineering and geophysical flows involving the transport of materials
such as cereals, rocks or sand. Many recent studies have focused on
the free-surface flow of a granular material 
\cite[][]{savage1991,forterre2008,bookpouliquen}, with particular
interest in steady flows down inclines \cite[see
e.g.][]{pouliquen1999b,pouliquen1999,mcelwaine2012:chute} or in the
collapse of a granular column
\cite[][]{lajeunesse2004,balmforth2005,lacaze2009,lagree2011}.  In
conjunction with studies of granular flow in shear cells and chutes,
this work has significantly advanced our understanding and modelling
of the dynamics of granular media. Nevertheless, a general theory
continues to be elusive, and it remains essential to consider
different configurations for granular flow that are readily set up in
the laboratory and explored theoretically.

In this paper, we explore the dense granular flow generated by
bulldozing a pile of grains over a level horizontal surface. The
problem can be conveniently set up in the laboratory by depositing the
pile on a table rotating underneath a stationary plate (with the
bottom of the blade held at the same height as the underlying
surface). The use of this rotating arrangement, instead of a blade in
rectilinear motion, allows the system to be recirculated so that the
dynamics can be observed over relatively long times. In addition, the
variation of the velocity with radial position along the blade
allows for a richer dynamics at the same time as enabling
one to study a range of bulldozing speeds all within a single
experiment. One of our aims is to provide a first experimental study
of this `rotating bulldozer' and to determine the key features of the
dynamics. With this in mind, we characterize the motion and shape of
the dune built up against the blade for a range of rotation rates,
dune volumes and initial positions, and for a number of different
granular media, ranging from glass ballotini to coarse sand and grit.

A second aim is to complement the experiments with theory.  For this
task, and following many conventional approaches to granular flow
problems with a free surface (Forterre \& Pouliquen 2008; Andreotti,
Forterre \& Pouliquen 2013), we consider a relatively crude
depth-averaged model that treats the granular medium as a continuum.
Such an approach has several limitations, not least of which are the
requirement that the flow be shallow and the need to incorporate a
prescription for the frictional internal stresses.  We therefore
supplement the depth-averaged model with simulations at the particle
level using the discrete element method (DEM), which is currently the
method of choice for many complex granular flows
\cite[][]{cundall1979,jim2009}.  DEM simulations offer a powerful
insight into granular flow dynamics, as idealized computations can be
performed without the complications associated with laboratory
experiments, and all quantities can be observed non-intrusively.  The
main drawbacks lie in the unrealistic contact laws they employ and the
limited number and geometric description of the particles. Despite
these drawbacks, satisfying agreement has been obtained with a number
of existing granular experiments.

Surprisingly, there are relatively few previous laboratory studies
of granular bulldozers, despite the classical early work by
\cite{bagnold1966} who considered the two-dimensional bulldozing of a
layer of sand of constant thickness. Because the bulldozer was held
below the sand surface and pulled at a fixed force, the dune volume
increased with time and episodic avalanching caused the bulldozer to
advance unsteadily.  Bagnold's laboratory experiments provided a
qualitative picture of the shape and the flow in the dune. However,
the experiments could not reach any steady state and no quantitative
measurements of the shape of the dune or the flow velocity were
provided.  More recently, experimental studies of `singing' or
`booming' sand \cite[][]{douady2006,andreotti2012} have used rotating
granular bulldozers to create avalanches. However, the purpose of
these experiments was to explore sound emission by the granular flow,
and the dynamical features of the flow itself were not the main
concern and were not therefore documented in detail.

The dynamics of granular bulldozers is also closely connected with the
problem of washboard patterns on dirt roads. Here, a plough or wheel
is driven over a granular surface; the dynamical interplay between the
bulldozed dune and the plough allows for an instability when the wheel
or the blade is free to move vertically, the oscillations of the plough
subsequently imprinting the washboard pattern
\cite[][]{mathur,taberlet2007,bitbol2009,percier2011,hewitt2012,percier2013}.
In order to provide lift, however, the surface of the plough is
inclined, a feature that, together with the vertical motion of the
plough, sets the washboard experiments apart from those that we document
here (where the plough is a fixed vertical blade).  Other recent,
related experiments have studied the drag on a plate pulled through a
granular medium
\cite[][]{geng2005,gravish2010,ding2011,guo2012,guillard2013}; these
studies were interested primarily in the drag force on the plate, and
again not the flow within the granular medium

We organize the paper as follows. In section 2 we describe in detail
the experimental apparatus; section 3 summarizes the results.  A
description of the depth-averaged modelling and discrete element
simulations appears in section 4.  Finally, in section 5, we discuss
our results and draw some conclusions, in particular highlighting open
questions and avenues for future research.

\section{Experimental set-up and procedure}

\subsection{The rotating bulldozer}

\begin{figure}
  \begin{center}
\includegraphics[width=\textwidth]{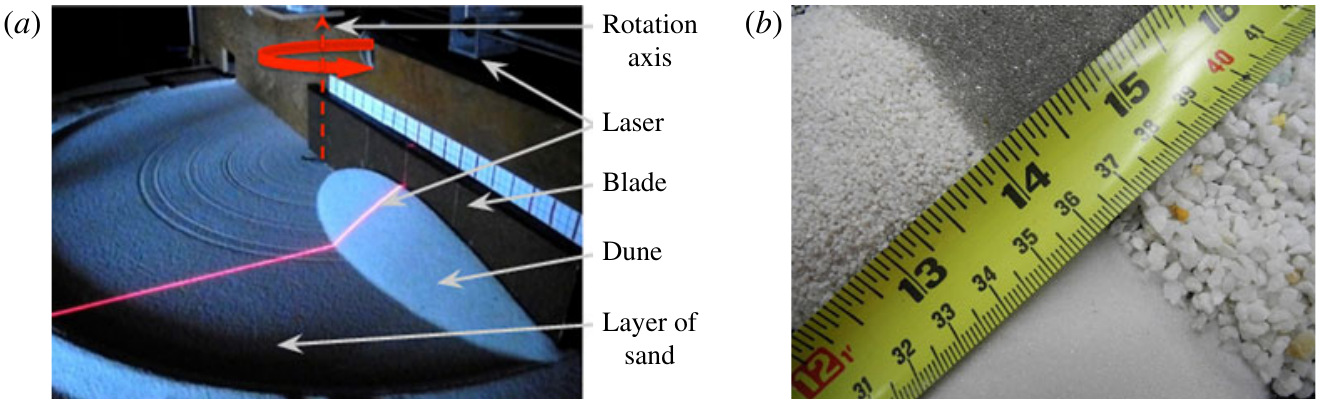}
    \caption{(a) Photo of the experimental setup. (b) Photograph of
      the four granular materials. From left and clockwise: aquarium
      sand ($d=0.9$ mm); spherical glass beads ($d=1$ mm); coarse grit
      ($d=3$ mm); and fine glass beads ($d=0.1$ mm).}
    \label{setup}
  \end{center}
\end{figure}

The experimental apparatus, shown in figure~\ref{setup}a, consisted of
a rotating table of 2.2\,m diameter above which a blade was fixed in
the laboratory frame. The table could be rotated at a fixed rate,
$\Omega$, in the range of 0.05--2\,rad\,s$^{-1}$, to a precision of 
a fraction of a percent.  We used a wooden
beam spanning the table for the bulldozer blade.  Both the beam and
the table were coated with sandpaper to prevent particles from sliding
over their surfaces.  We orientated the blade vertically at a fixed
height of $15\pm 2$\,mm above the table.

To begin each experiment, we first poured the granular material onto the
rotating table, filling up the gap underneath the blade and ploughing
out a layer of constant thickness and compaction. The vertical
separation of the blade from the bed avoided particles becoming jammed
and crushed between the blade and the table.  Although the thickness of
this pre-existing layer adds to the parameters of the problem, we ran
additional experiments to check that variations in its depth over the
range 10--30\,mm did not significantly change the observed
phenomenology at the rotation rate, $\Omega=0.05$\,rad\,s$^{-1}$.
The DEM simulations in \S 4.3 offer further gauging of the effect of
the underlying layer.

Once the pre-existing layer was in place, we stopped the table, and
then slowly poured an additional mass of grains onto the layer at a
selected position.  This created an initial pile in the form of a
nearly conical mound of given mass $m$, with a slope given by the
static angle of repose of the particular granular material and centred
at some initial radius $r_0$ from the centre of the table.  For the
present study, we used $r_0=15$, $25$ or $35$\,cm, and masses of
$m=250$\,g, $500$\,g, $750$\,g and $1000$\,g.  The table was then set in
motion again, accelerating to a prescribed rotation rate, $\Omega$,
with equivalent rotation period $T_{rot}=2 \pi/\Omega$, in typically
less than 2 seconds (well before the pile hits the blade).  The
collision of the pile and the subsequent dynamics then took place
at constant rotation rate.  

\subsection{Granular materials}
In this study, we considered four different granular media,
two of irregular shape and two spherical, as shown in figure~\ref{setup}b: 
\begin{enumerate}[i) ]
\item aquarium sand (left; manufacturer PetCo);
\item coarse grit (right); 
\item $1$ mm diameter glass beads (top; A-100 from Potters Industries);
\item $0.1$ mm diameter glass beads (bottom; \#8 from Kramer Industries).
\end{enumerate}
\noindent The mean particle size, or their ranges, for all of the granular
materials are summarized in table \ref{tableau_1}.

Granular materials are often characterized by static and dynamic
angles of repose, although these have no universally agreed definitions.
To estimate these for each of our four materials, we half-filled a
drum of inside diameter $138$\,mm and axial length $93$\,cm.  Rotating
the drum at low speeds led to episodic avalanching punctuated by
periods of rigid body rotation (the so-called ``slumping'' regime;
\textit{e.g.}  \cite{gdr2004}). By fitting a line to the free surface, we
then measured the average angle at which avalanches started,
$\theta_{start}$, and ended, $\theta_{stop}$. Rotation of the drum at
faster speeds eventually precipitated continuous avalanching (the
``rolling'' regime) at a dynamic angle of repose, $\theta_{c}$.  For
the glass ballotini and sand, the free surface remained fairly linear
even at the higher rotation rates.  However, the surface of the grit
was significantly nonlinear during continuous avalanching, indicating
that $\theta_c$ only roughly characterizes the slope of the entire
surface for this material.  These angles are listed in table
\ref{tableau_1} for the four granular materials, along with the
experimentally observed angle of the avalanching dune $\theta_M$
(discussed in more detail in section \ref{sec:theta_m}).  The data are
similar to cruder measurements that we made of the surface slopes of
slowly built up sandpiles \cite[{\textit{cf.}}][]{Borzsonyi2008}.

\begin{table}
  \begin{center}
    \begin{tabular}{ l c c c c c c}
     &  $d$ / mm  & $\rho$ / g cm$^{-3}$
      & $ \qquad \theta_{start} \qquad $ & $\qquad \theta_{stop} \qquad $ 
      & $\theta_c$ & $\theta_M$ \\        
      Ballotini & $1 \pm 0.2$  & $1.48-1.61$ & $24.9 \pm 0.7 \!\char23$  
      & $22.9 \pm 0.5 \!\char23$ & $23.7 \pm 0.6\!\char23$  & $23.4\pm0.5\!^\char23$\\ 
      Ballotini  & $0.12 \pm 0.03$ & $1.50-1.60$ & $26.6 \pm 0.7 \!\char23$ 
      & $23.8 \pm 0.6 \!\char23$&  $25.3 \pm 0.5 \!\char23$  & $23.5\pm0.5\!^\char23$\\
      Aquarium sand & $0.9 \pm 0.15$ & $1.50-1.67$ & $37.9 \pm 1.3 \!\char23$ 
      & $33.7 \pm 1.1 \!\char23$ & $36.1 \pm 1.0 \!\char23$  & $36.1\pm0.5\!^\char23$\\
      Coarse grit & $3 \pm 0.9$  & $1.46-1.64$ & $39.2 \pm 0.9 \!\char23$ 
      & $33.1 \pm 0.8 \!\char23$ &  $36.3 \pm 1.5 \!\char23$ & $35.5\pm0.5\!^\char23$\\
    \end{tabular}
  \end{center}
  \caption{Mean particle diameter $d$, apparent density $\rho$,
and characteristic angles
    for the four different granular media. 
The mean diameter of the aquarium sand and coarse grit particles have
been estimated by direct visualization and fitting a Gaussian distribution
to the recorded particles sizes; the quoted error is the standard 
deviation \cite[see][for more
details]{sauret2012}. For the ballotini, we quote the particle range
and its median as provided by the producers.
The apparent densities are determined by measuring the mass of one litre
of each of the materials with either loose packing or after compaction.
$\theta_{start}$ and $\theta_{stop}$
    denote the starting and stopping angles for avalanches in a slowly
    rotating drum;
    $\theta_{c}$ is the the dynamic avalanching angle at the initiation of the
    continuous flow regime in the same drum rotated at higher speed.
    $\theta_M$
    is the angle of the low-rotation-rate master curves observed in the bulldozing
    experiments, as described in more detail in section \ref{sec:theta_m}.}
  \label{tableau_1}
\end{table}
  
\subsection{Diagnostic method: calibration and topography of the dune}
\label{measurement_annexe}

\begin{figure}
  \begin{center}
\includegraphics[width=0.95\textwidth]{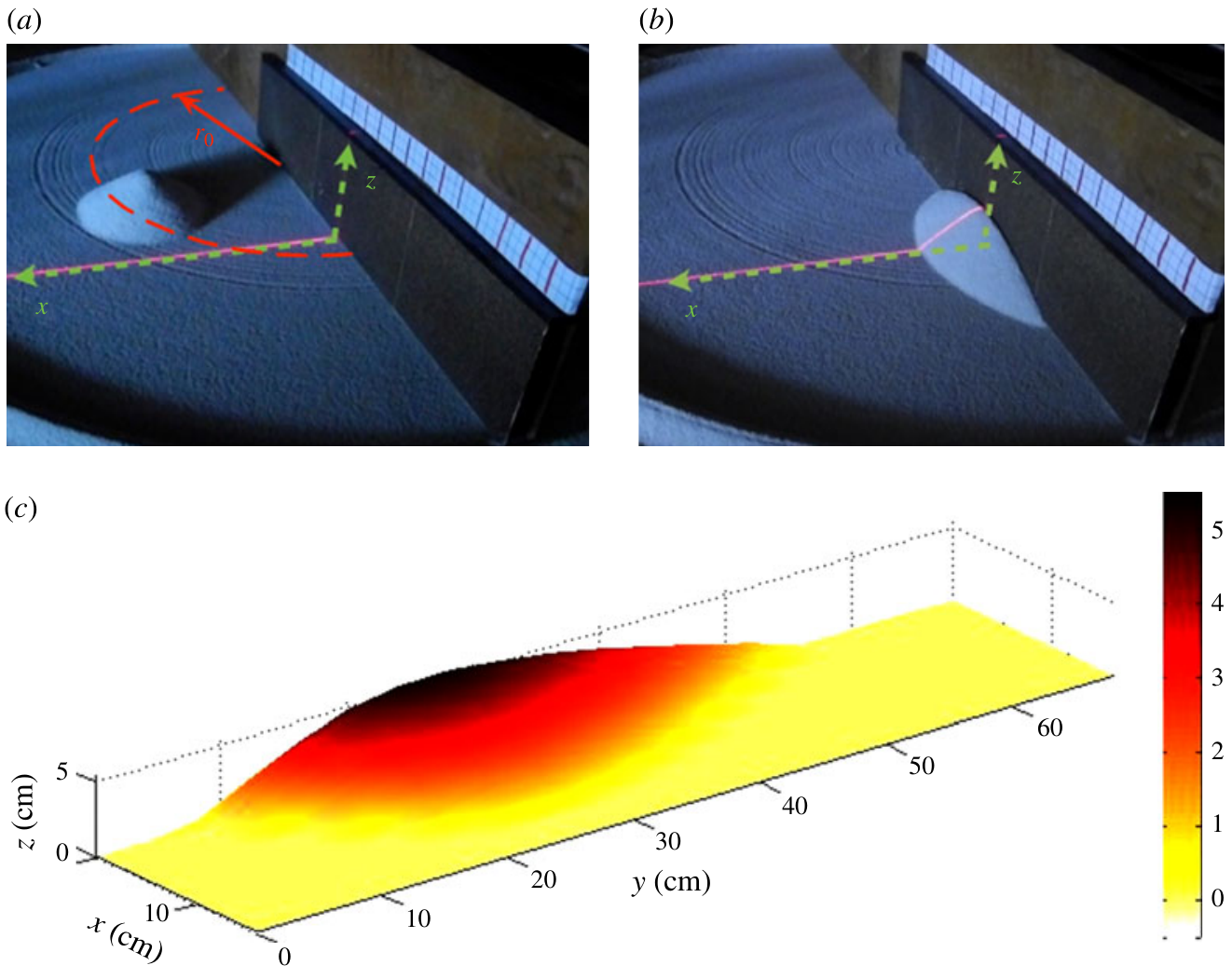}
    \caption{Coordinate system and laser measurement technique
      used to characterize the surface of the bulldozed pile: a
      Cartesian coordinate system is
      orientated so that the $z-$axis points vertically upwards, $x$
      is perpendicular to the bulldozer's blade and
      $y$ runs laterally along its front face.  (a) Without 
      topography, the laser line (red) runs along the top of the
      uniform underlying layer and defines a section in $x$ at $z=0$
      for a particular position in $y$; the $(x,z)$ plane at this location in $y$ is shown. The dashed
      semi-circle indicates the radial position $r_0$ of the centre of
      the initial pile. (b) When the
      topography of the dune cuts
      through the line of the laser, the deviation from the original
      straight line can be used to measure the elevation of the
      surface, $z=h(x,y,t)$.  (c) The reconstructed
      dune profile for a mass $m=1000$ g of $1$ mm glass
      beads, with rotation rate $\Omega=0.05$ rad\,s$^{-1}$ and
      initial position $r_0=25$\,cm. 
        }
    \label{method_topo}
  \end{center}
\end{figure}

To describe the geometry of the bulldozed dune, we use the Cartesian
coordinate system $(x,y,z)$ illustrated in figure~\ref{method_topo}.
The axes are orientated so that the $z-$axis points vertically upwards
with $z=0$ corresponding to the base of the bulldozer (or the surface
of the underlying granular layer), the $x-$axis is perpendicular to
the bulldozer blade and $y$ runs along its front face.
The surface profile of the dune, $z=h(x,y,t)$, is equivalent to
the local depth of material above the pre-existing granular layer.
Note that, because of
the finite thickness of the blade and the way that we positioned the
wooden beam, the origin of the coordinate system does not coincide
with the rotation axis of the table. Instead, the origin is offset
from the centre of the rotating table and is positioned at the closest
point along the blade.

To determine the profile of a dune, we projected a laser line onto its
surface, measuring the deflection away from the pre-existing layer due
to that topography as shown in figure~\ref{method_topo}a,b
\cite[{\textit{cf.}}][]{pouliquen1999b}.  The laser line was orientated
perpendicular to the blade, and fixed on a traverse such that the
laser could be swept along the length of the blade. The perpendicular
profile ({\it i.e.}  in an $(x,z)$-plane) of the dune could then be
measured at various lateral (radial) positions along the blade.
However, most of our perpendicular profile measurements were taken at the
lateral location, $y_0\approx r_0$ corresponding to the position where the
centre of the initial pile hits the blade.  To deal with effects of
perspective and projection, we first performed a calibration using
images of square boards, suitably positioned in the field of
view of the camera. (See \cite{sauret2012} for more details.)  This exercise
furnished the proper map between the deflections measured in the
pixels of the camera and the actual distances in three-dimensional space.
Figure~\ref{method_topo}c shows a typical example of the
reconstruction of the entire topography of a dune.

\section{Experimental results}

\subsection{Phenomenology}

\begin{figure}
  \begin{center}
    \includegraphics[width=12cm]{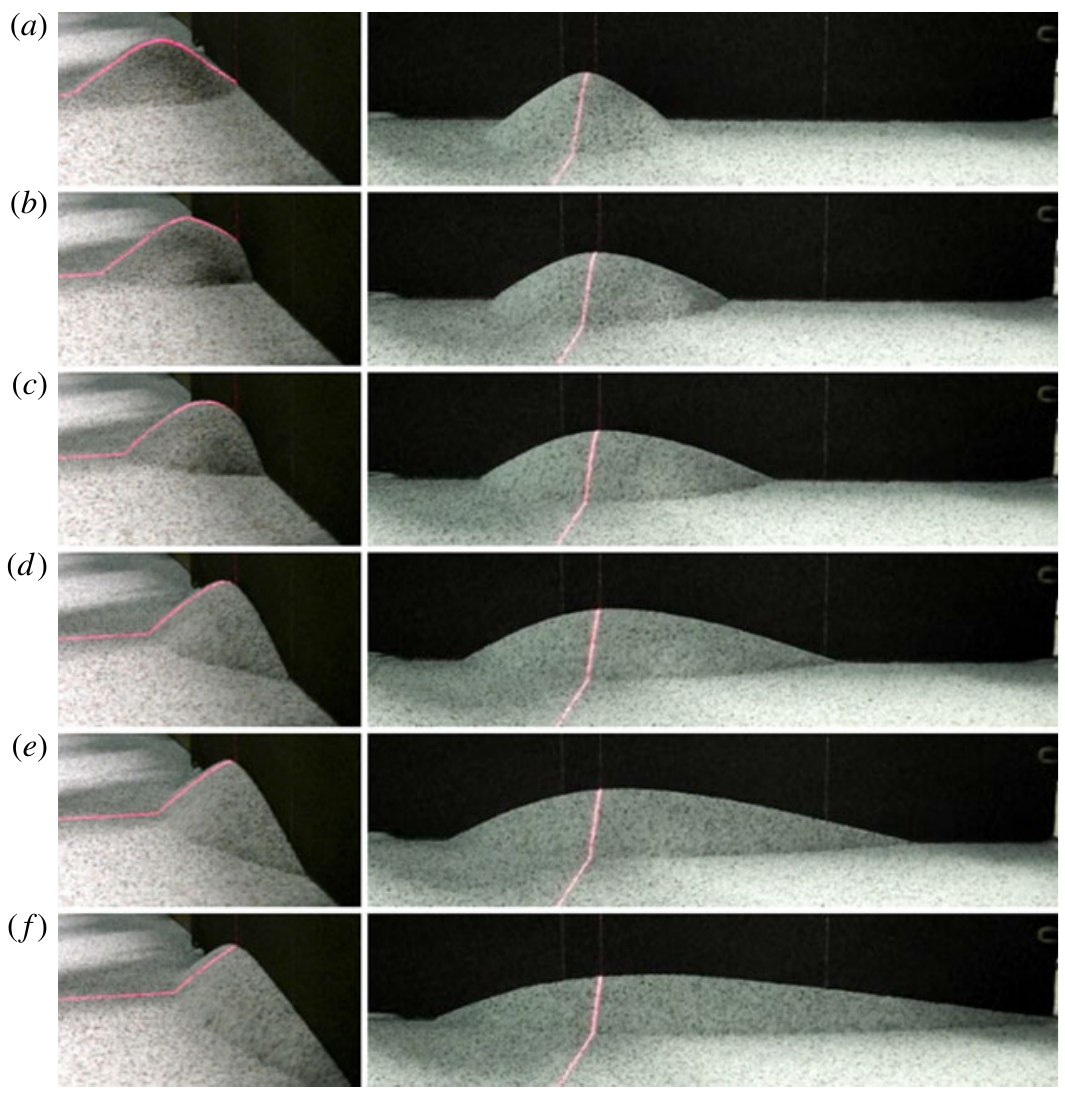}
    \caption{Successive pictures of the bulldozed dune
      from a side view (left) and a front view (right) for $r_0=25$
     \,cm, $\Omega=0.05$\,rad\,s$^{-1}$ and $m=1000$ g. The times of the
      snapshots are (a) $t/T_{rot}=0$, (b) $0.032$, (c) $0.064$, (d)
      $0.127$, (e) $0.24$ and (f) $0.64$.  In all pictures, the 
      laser line highlights the topography of the dune at the 
      location $y_0$. Videos of these experiments are available as
      supplementary materials.  }
    \label{fig:phenomenology}
  \end{center}
\end{figure}

\begin{figure}
  \begin{center}
\includegraphics[width=13cm]{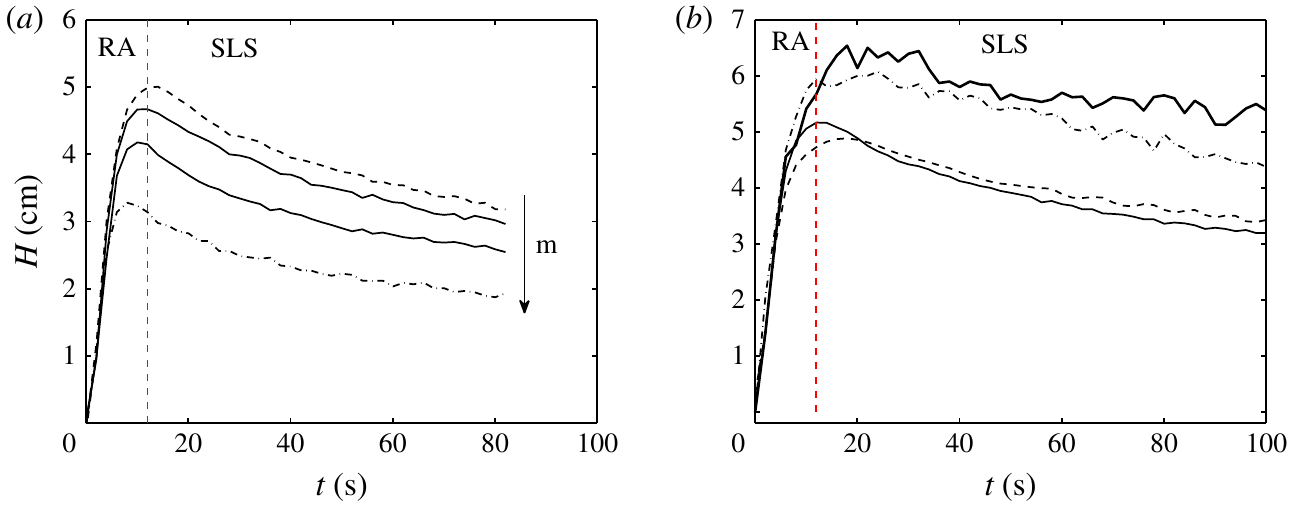}
    \caption{Time series of the depth of the dune against the blade,
      $h(0,y_0,t)\equiv H(y_0,t)$ at $y=y_0\approx r_0=25$\,cm, with
      $\Omega=0.05$\,rad\,s$^{-1}$, for (a) $1$ mm glass beads with
      varying mass ($m=250$, 500, 750 and 1000 g)
      with the arrow indicating decreasing initial mass; (b)
      the four granular materials with fixed mass $m=1000$ g (grit -
      thick line, aquarium sand - dashed-dotted line, $1$ mm glass
      beads - dashed line, $0.1$ mm glass beads - solid line).  The
      vertical lines at $t=12$\,s indicate the rough switch over
      from the rapid initial adjustment (labelled ``RA'') to the
      slower lateral spreading phase (``SLS'').  }
    \label{fig:profile_syst_mass_what_21}
  \end{center}
\end{figure}

A collision between a conical pile of aquarium sand and the blade is
illustrated in figure~\ref{fig:phenomenology} and documented further
in the videos available as supplementary material. The bulldozing
dynamics takes place in two phases. The collision
prompts a rapid initial rearrangement of the granular material into
an avalanching dune driven primarily {perpendicular} to the blade (figure
\ref{fig:phenomenology}b,c).  This avalanching state becomes
quasi-steady in its perpendicular motion, but subsequently spreads
laterally over a longer time scale (figure~\ref{fig:phenomenology}d-f).  The two phases are evident in figure
\ref{fig:profile_syst_mass_what_21}, which plots time series of the
depth of the dune at the position $y_0$ along the blade
corresponding to the initial radius of the pile, i.e.
$h(0,y=r_0,t)\equiv H(y=r_0,t)$, for a number of different
experiments.  Note that we fix the origin of time, $t=0$, as the
moment when the edge of the initial pile makes contact with the
blade of the bulldozer.

The rapid adjustment phase spans a time $T_i$ of the order of tens of
seconds, or a fraction of the rotation period. This time scale can be
rationalized in terms of the characteristic length of time required
to reorganize a pile of initial radius $R_0$ by
flow with speed $U=r_0\,\Omega$,
$T_i\sim R_0/(r_0\,\Omega)\approx12\,$s
(for $R_0\approx 15$\,cm, $r_0\approx 25$\,cm and 
$\Omega=0.05$\,rad\,s$^{-1}$),
which is indicated in figure
\ref{fig:profile_syst_mass_what_21}a,b.
Lateral spreading, on the other hand, takes
place over a time scale $T_q$ spanning multiple rotations, suggesting
that $T_q=O(T_{rot})$, which is consistent with the predictions of the
depth-averaged model presented below in section \ref{sec:lateraldata}.

\subsection{Dynamics perpendicular to the blade}\label{sec:theta_m}

\begin{figure}
  \begin{center}
    \includegraphics[width=8.5cm]{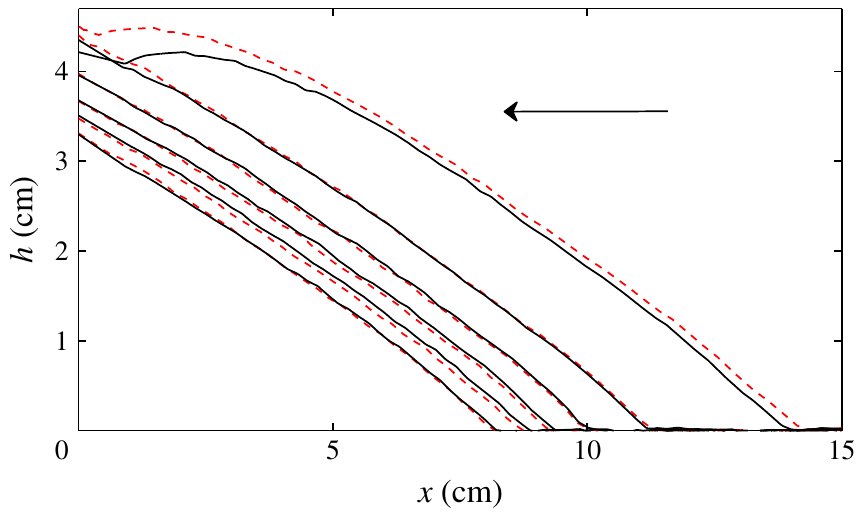}
    \caption{Profiles of two bulldozed dunes at $y=25$\,cm for initial
      piles of $m=1$ kg of $1$ mm glass beads located at $r_0=25$
     \,cm, with $\Omega=0.05$\,rad\,s$^{-1}$. The profiles are taken
      every 10 seconds starting at $15$ seconds after the pile
      hits the blade, with the direction of increasing time being
      shown by the arrow. The profiles for the two distinct
      experiments are shown by dashed and solid lines.
    }
    \label{fig:reproducibility_vir}
  \end{center}
\end{figure}

We first consider the quasi-steady perpendicular profiles of the
avalanching dunes at fixed lateral position $y$.  Profiles from two
different experiments with the aquarium sand and similar initial
conditions and rotation rate ($r_0=25$\,cm, $m=1$ kg and $\Omega=0.05$\,rad\,s$^{-1}$) are shown in figure~\ref{fig:reproducibility_vir}. The
two earliest profiles show slight variations due to minor differences
in the initial piles: although the mass is fixed, particles can be
packed slightly differently \cite[see e.g.][]{jaeger1996} and the
exact position of the initial pile can vary by a few millimetres.
These variations become less significant during the initial rapid
adjustment into the quasi-steady bulldozed dune, leaving profiles that
agree to within the measurement errors of the laser system, which are
approximately 2\,mm. The profiles then slowly reduce in height over
the time scale $T_q$ due to lateral spreading. Thus, the evolving
topography of the dune is reproducible in the experiments.

\begin{figure}
  \begin{center}
\includegraphics[width=13.5cm]{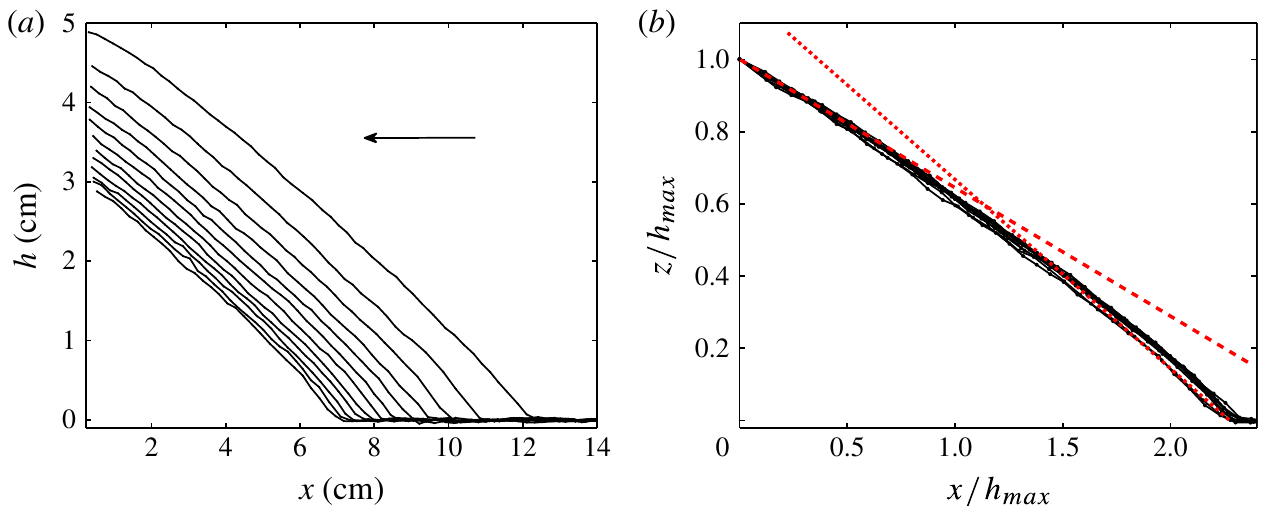}
    \caption{ Perpendicular profiles at $y=y_0$ 
for a bulldozed dune of $1$\,mm glass
      beads with $\Omega=0.05$\,rad\,s$^{-1}$, $y_0\approx r_0=25$\,cm
      and $m=1000$ g. {Panel (a) shows} plots dimensional dune profiles,
      $h(x,y_0,t)$, against $x$; the profiles decay with time and are
      measured every $8$ seconds. {Panel (b) shows} the scaled profiles,
      $h(x,y_0,t)/H(y_0,t)$ against $x/H(y_0,t)$, where
      $H(y,t)=h(x=0,y,t)$ is the depth against the blade.
      The dotted line shows the local slope at
        $x=0$, and the dashed line is the local slope at the flow
        front (as $h \rightarrow 0$).}
    \label{fig:profile_syst_mass_a1}
  \end{center}
\end{figure}

We plot quasi-steady profiles for $1$\,mm glass beads in figure
\ref{fig:profile_syst_mass_a1}a for an initial mass $m=1000$\,g.  The
slope of the dune remains nearly constant over the duration of the
period shown, even though the perpendicular area falls by a factor of
about two due to lateral spreading.  Interestingly, when we scale
$h(x,y,t)$ and $x$ by the depth at the blade,
$h(x=0,y,t)\equiv H(y,t)$, all the profiles collapse onto a common
curve, as shown in figure~\ref{fig:profile_syst_mass_a1}b.  To leading
order, this ``master curve'' is linear, but flattens out slightly at
the blade as $x \rightarrow 0$ and steepens up near the front of the
dune as $h \rightarrow 0$.

\begin{figure}
  \begin{center}
    \includegraphics[width=\textwidth]{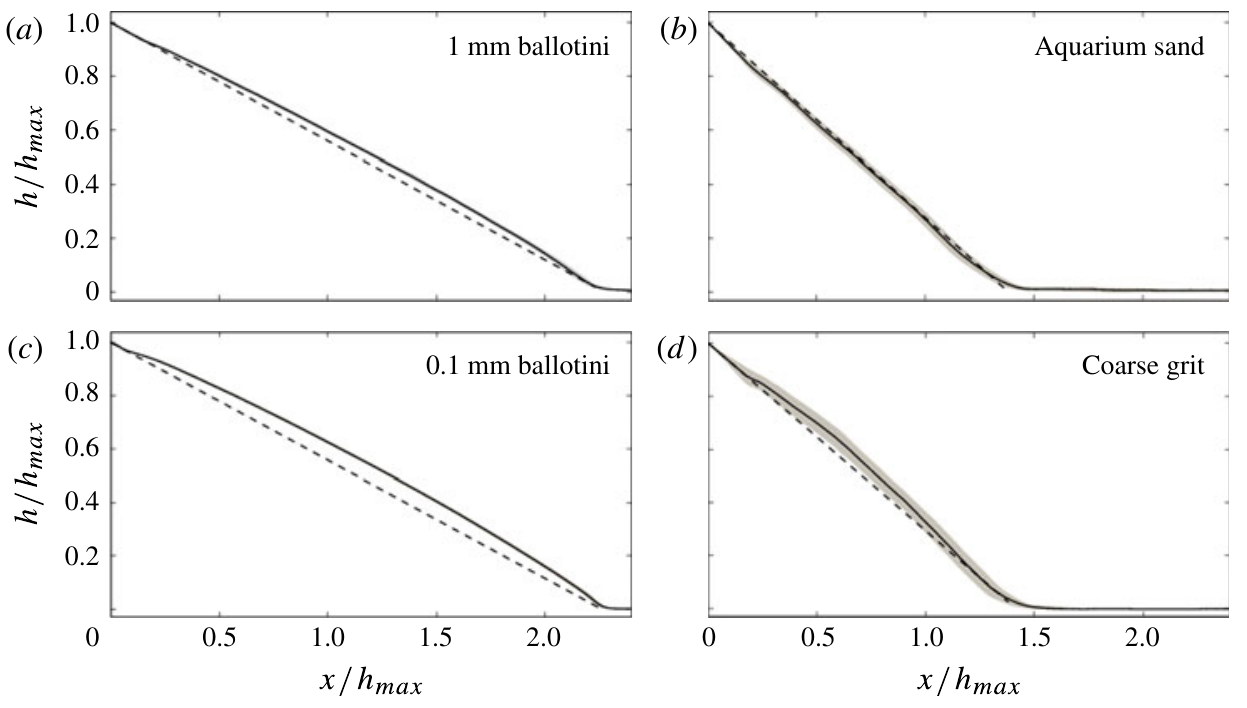}
    \caption{ Master curves for: (a) $1$ mm glass beads; (b) aquarium
      sand; (c) $0.1$ mm glass beads; (d) coarse grit {with},
      $\Omega=0.05$\,rad\,s$^{-1}$, $r_0=25$\,cm and $m=1000$ g.  The
      master curves are constructed by averaging between 10 and
      15 scaled perpendicular profiles at $y=y_0\approx r_0$ (see figure
      \ref{fig:profile_syst_mass_a1}) taken every 4 s.  The grey
      band shows the standard deviation of the scaled profiles about
      the mean master curve.  The dashed line linearly interpolates
      between the blade and the tip of the dune.  }
    \label{fig:profile_syst_mass_what}
  \end{center}
\end{figure}

Figure~\ref{fig:profile_syst_mass_what} shows master curves
(constructed by averaging between ten and fifteen profiles taken every
four seconds after the dune converges to its quasi-steady state) for
our four granular materials with $\Omega=0.05$\,rad\,s$^{-1}$,
$r_0=25$\,cm and $m=1000$\,g.  In the figure, the grey band indicates
the standard deviation of the profiles about the average.  For the
glass beads, the profiles all collapse closely to the master curve
(the corresponding grey bands are relatively thin), while there are
more significant variations about the average for the irregularly
shaped sand and grit particles.  The dashed lines in the figure
indicate the leading-order slopes of the master curves, obtained by
linearly interpolating between the intersection of the master curve
with the blade and the front of the dune.  The angles corresponding to
these slopes, which we denote by $\theta_M$, are similar to the
dynamic avalanching angles $\theta_c$ measured in the rotating drum,
as reported in table \ref{tableau_1}. Although the master curves are
mostly linear, there is some curvature to these profiles, particularly
for the glass ballotini, suggesting an interesting effect of particle
shape.

\begin{figure}
  \begin{center}
    \includegraphics[width=\textwidth]{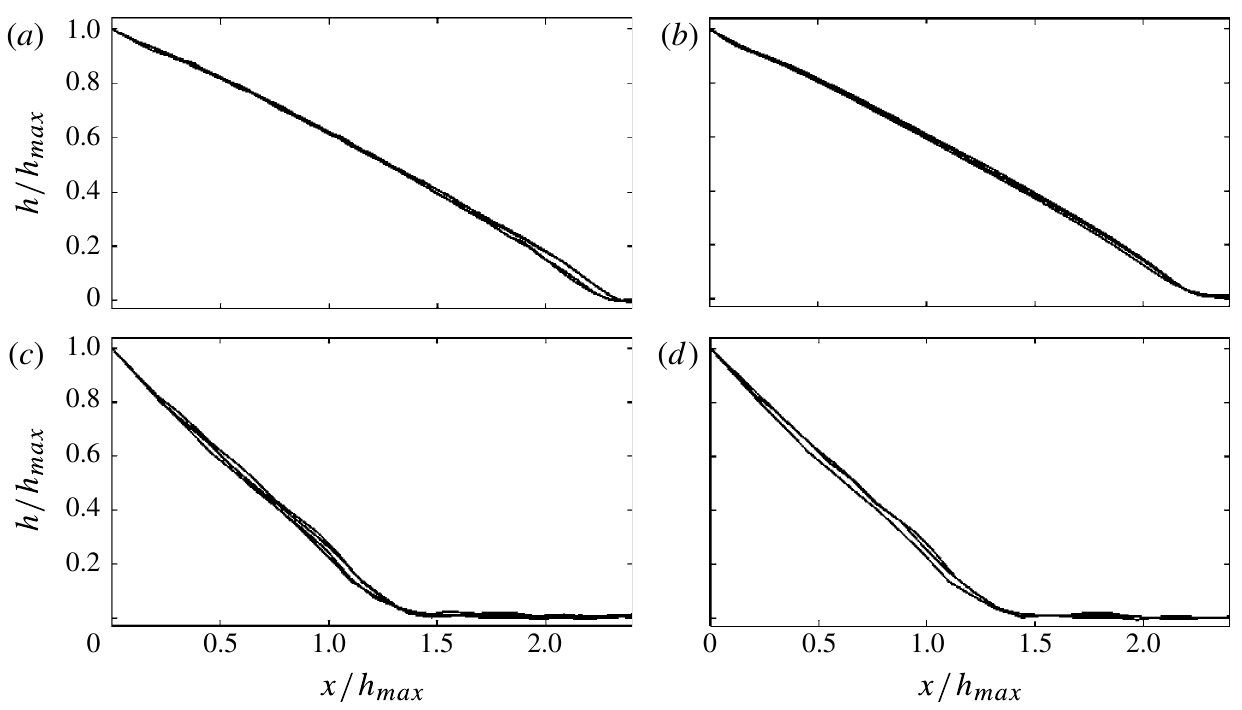}
    \caption{
      Master curves {with $\Omega=0.05$\,rad\,s$^{-1}$} at $y=y_0\approx r_0$ for: (a) $r_0=25$\,cm 
and various initial mass
      ($m=250$, 500, 750, 1000 g) for $1$ mm glass beads; (b) $m=1000$
      g and varying initial position ($r_0=15$, 25 and 35\,cm) for
      $1$\,mm glass beads; (c) $r_0=25$\,cm and various initial mass
      ($m=250$, 500, 750, 1000 g) for aquarium sand; (d) $m=1000$ g and
      varying initial position ($r_0=15$, 25 and 35\,cm) for aquarium
      sand.  }
    \label{fig:reproducibility_sl}
  \end{center}
\end{figure}

\begin{figure}
  \begin{center}
    \includegraphics[width=\textwidth]{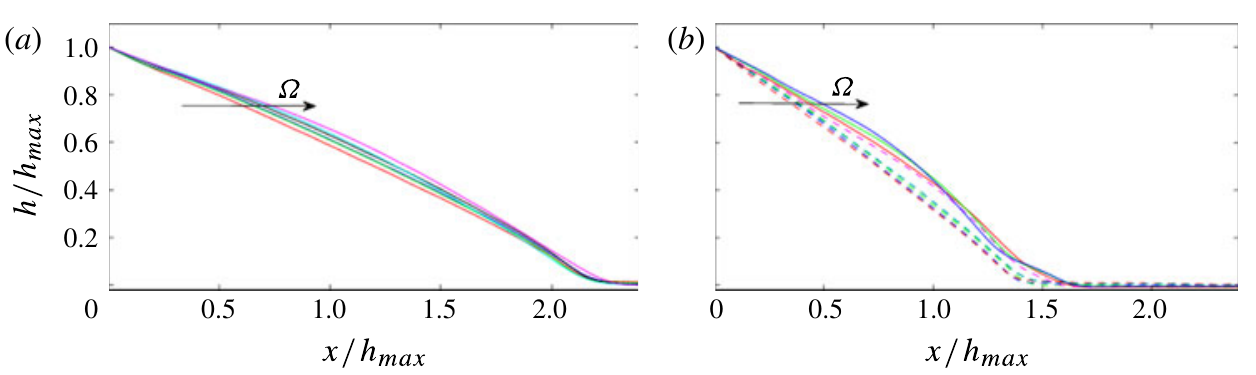}
    \caption{Master curves {with $m=1000$ g} at $y=y_0\approx r_0=25$cm and
different rotation rate, for (a) $1$ mm
      glass beads ($\Omega=0.05$, $0.1$, $0.2$, $0.4$, $0.8$ and 
      1\,rad\,s$^{-1}$); (b) aquarium sand 
($\Omega=0.05$, $0.2$, $0.4$\,rad\,s$^{-1}$ - dashed, and
 $\Omega=0.8$, $1$, $1.5$, $2$\,rad\,s$^{-1}$ - solid). 
The arrows indicates the direction of increasing $\Omega$.  }
    \label{fig:increase_F}
  \end{center}
\end{figure}

In figures \ref{fig:reproducibility_sl}a--d, we plot master curves
constructed from experiments with varying initial mass $m$ and radial
position $r_0$, for both 1\,mm glass beads and aquarium sand. For each
material, the various experiments collapse onto the same master curve.
Similarly, when we take measurements of the perpendicular profiles of
the bulldozed dunes at different radial positions $y$ along the blade,
we also recover a common master curve.  Only when we significantly
speed up the rotation of the table do we observe any change in the
characteristic structure of the master curves. As illustrated in
figures \ref{fig:increase_F}a,b, when the angular velocity $\Omega$ is
increased, the curves become more nonlinear, somewhat similarly to the
fashion in which the surface of flow in granular drums develops a
characteristic `S-shape' \cite[][]{gdr2004,taberlet2006}.

More generally, one would expect that the speed dependence of the master
curves would arise through the local Froude number, $\Fr = |U_b| / \sqrt{g
  H} \equiv y \Omega / \sqrt{gH}$, where $|U_b|$ is the speed of the
bulldozer. The experiments in figures
\ref{fig:profile_syst_mass_a1}-\ref{fig:reproducibility_sl} are
conducted at relatively low Froude number, $\Fr=O(10^{-2})$ or less,
and so the dependence on the radial position, $y$, and the depth at the blade,
$H$, is not evident; the curves in these figures characterize the
zero-Froude-number limit.  Only when the rotation rate is increased by
two orders of magnitude, as in figure~\ref{fig:increase_F}, does the
Froude-number dependence become apparent. In any event, we conclude
that the perpendicular {profiles} collapse to Froude-number-dependent
master curves that are characteristic of the granular material,
much as the shape of the free surface reflects the flow dynamics in a
rotating granular drum. 

A final noteworthy point is that the
perpendicular avalanching was always observed to be steady,
except over the longer time scale of
lateral spreading. This contrasts with what one observes in
rotating drums, wherein flows become intermittent at low
rotation rates and occur {\it via} episodic avalanching.
In fact, our DEM simulations did show intermittent motion at low
bulldozing speed (see \S 4.3), with avalanching becoming 
temporally irregular, if not episodic. Thus, 
our experiments were probably not conducted
at sufficiently low speeds to observe unsteady
states.

\subsection{Lateral spreading of the dune}\label{sec:lateraldata}

\begin{figure}
  \begin{center}
    \includegraphics[width=11cm]{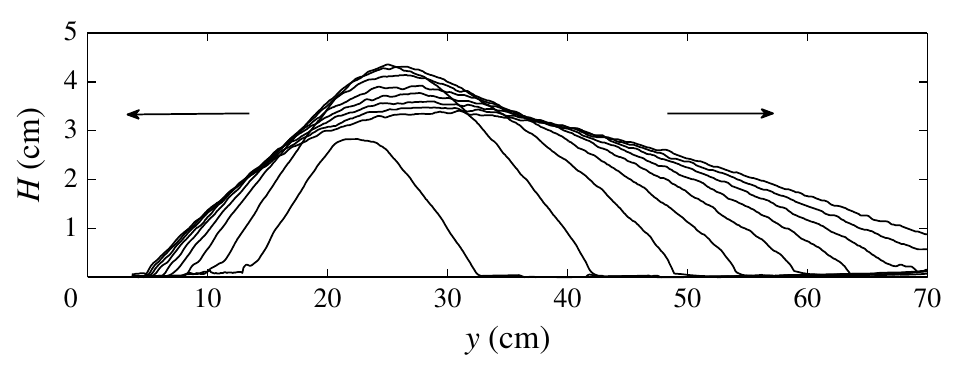}
    \caption{ Lateral profiles, $H(y,t)=h(x=0,y,t)$, for $1$mm glass
      beads with $\Omega=0.05$\,rad\,s$^{-1}$, $m=1000$ g, and initial
      position $r_0=25$\,cm.  The arrows indicate the direction of
      increasing time and the profiles are taken every $5$ seconds
after the dune hits the blade.}
    \label{fig:profile_syst_mass_what_2a11}
  \end{center}
\end{figure}

In addition to the quasi-steady perpendicular profile, the rearrangement
of the granular mound prompted by the collision of the initial
conical pile with the bulldozer also generates a distinctive lateral
profile, $H(y,t)=h(0,y,t)$, along the blade.  Figure
\ref{fig:profile_syst_mass_what_2a11} displays the spreading of this
lateral profile for 1 mm ballotini. Initially, the collision
establishes a shape that is almost mirror symmetric
in $y$ about the centre of the initial pile, $y=y_0\approx r_0$. The
dune subsequently spreads both inwards, towards the centre of the
rotating table, and outwards to larger radii. The rate of spreading
is, however, asymmetrical, with the dune spreading faster at larger
radii and the inward edge eventually coming to rest at a finite
distance from the centre of the table.  This asymmetry results
in part from the increase of the perpendicular velocity $\Ub=-\Omega\,y$
with radial position $y$. However, because the front of the blade is
positioned at a small distance $\delta$ in front of
the centre of the rotating table, there is an additional tangential
velocity along the blade, $\Vb=\Omega\,\delta$, which advects the dune
radially outwards ({\textit{cf.}} \S \ref{sec:lateralmodel}).

\begin{figure}
  \begin{center}
    \includegraphics[width=12cm]{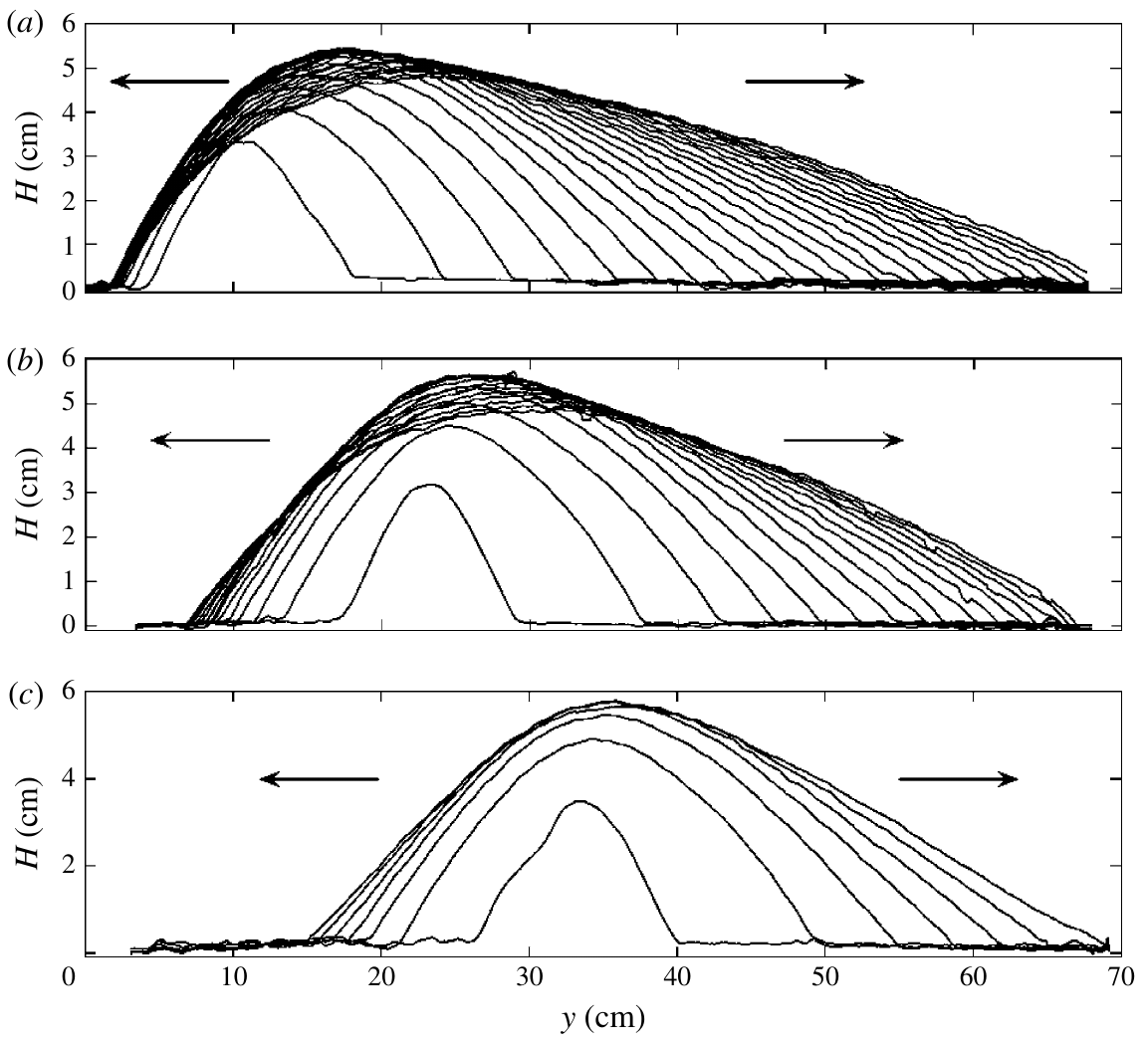}
    \caption{ Lateral profiles 
      $H(y,t)=h(x=0,y,t)$ for aquarium sand with
      $\Omega=0.05$\,rad\,s$^{-1}$, $m=1000$\,g, and varying initial
      position: (a) $r_0=15$\,cm, (b) $r_0=25$\,cm and (c) $r_0=35$\,cm.
      The arrows indicate the direction of increasing time and the
      profiles are taken every $5$ seconds after the dune hits the blade.}
    \label{fig:profile_syst_mass_what_211}
  \end{center}
\end{figure}

In figure~\ref{fig:profile_syst_mass_what_211}, we plot the evolution
of the lateral profile for dunes of aquarium sand for different initial
positions $r_0$; the lateral spreading depends significantly upon the
value of $r_0$.  In all three cases, the dune spreads in the same
asymmetrical fashion.  However, 
unlike for the perpendicular profiles discussed earlier, there appears
to be no straightforward scaling that collapses all these data.
Similarly, we are unable to collapse the
lateral profiles of dunes with different initial masses; only from a
more qualitative perspective are the different experiments comparable.
Likewise, lateral spreading is qualitatively, but not quantitatively,
similar for the other granular material ({\textit{cf.}} figure
\ref{fig:profile_syst_mass_what_2a11}). 

\begin{figure}
  \begin{center}
    \includegraphics[width=\textwidth]{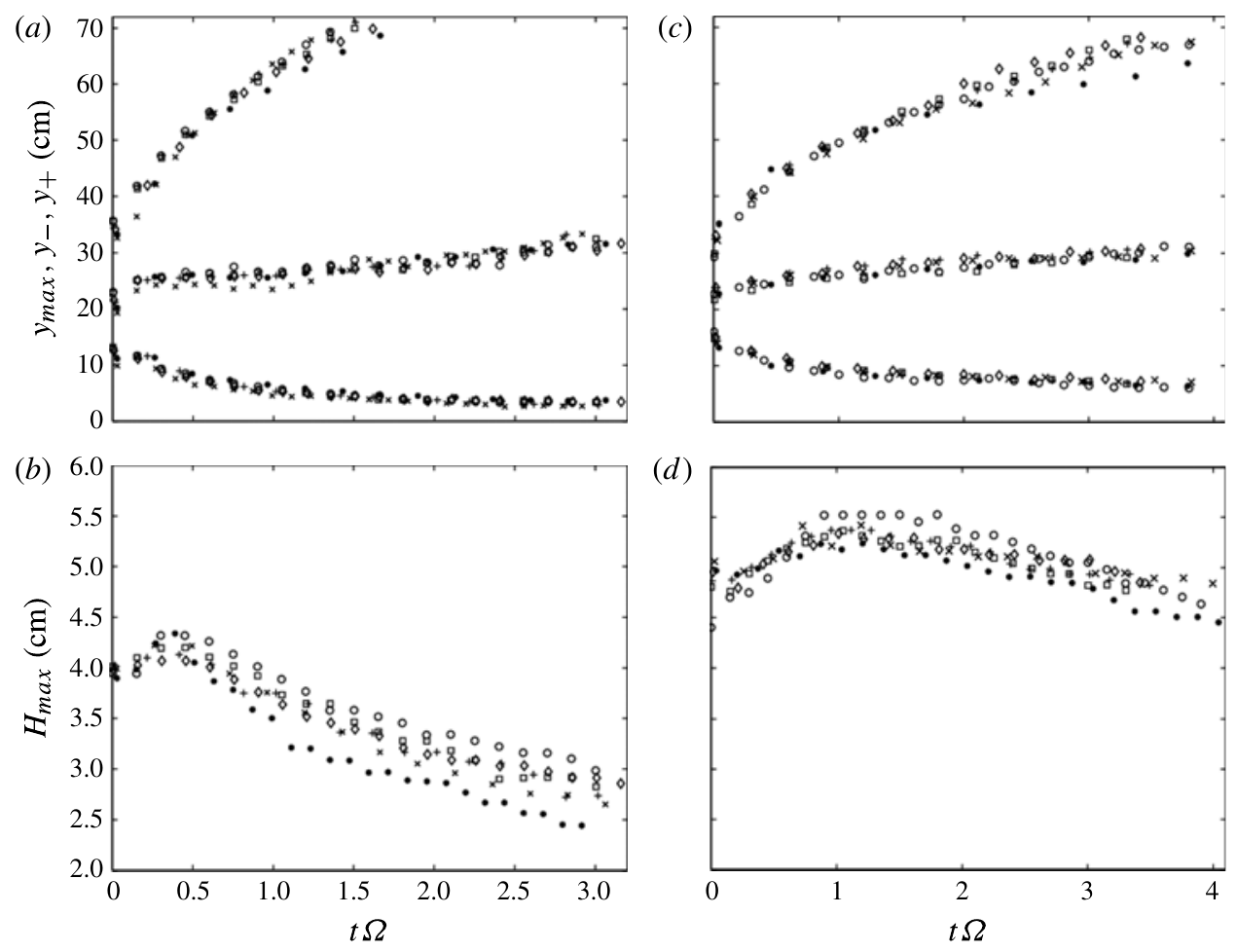}
    \caption{Time series of (a,c) the locations along the blade of
      the inner and outer edges of dunes, $y_-$ and $y_+$ and the
      position of maximum depth, $y_{max}$, and (b,d) the maximum
      depth, $H_{max}={\rm Max}_y\{H(y,t)\}$.  The results are plotted
      against $\Omega t$ for different rotation
      rates ($\Omega=0.05$\,rad\,s$^{-1}$ - $\circ$,
      $0.1$\,rad\,s$^{-1}$ - $\square$, $0.2$\,rad\,s$^{-1}$ - 
      $+$, $0.4$\,rad\,s$^{-1}$ - $\diamond$,
      $0.8$\,rad\,s$^{-1}$ - $\times$ and 1\,rad\,s$^{-1}$ -
      dots).  All the piles have $m=1000$ g and $r_0=25$cm.
      In (a,b), data for 1\,mm ballotini are shown;
      (c,d) show data for aquarium sand.
    }
    \label{Num_Theo_profile_b}
  \end{center}
\end{figure}

Nevertheless, data from experiments with 
the same initial mass and position but different rotation rate
can be conveniently collapsed, as illustrated in figure
\ref{Num_Theo_profile_b}. Time series of the locations of
the inner and outer edges, denoted by $y_-(t)$ and $y_+(t)$, 
are plotted for the 1\,mm glass beads and the aquarium sand.  Also shown is the
time series of the maximum depth of the dune against the blade,
$H_{max}(t)={\rm Max}_y\{H(y,t)\}$, along with the radial location
$y_{max}(t)$ at which that maximum occurs.  By plotting these
quantities against the dimensionless time $\Omega \, t$, the data 
collapse, confirming that the rotation rate controls the
spreading dynamics.

\begin{figure}
  \begin{center}
    \includegraphics[width=13cm]{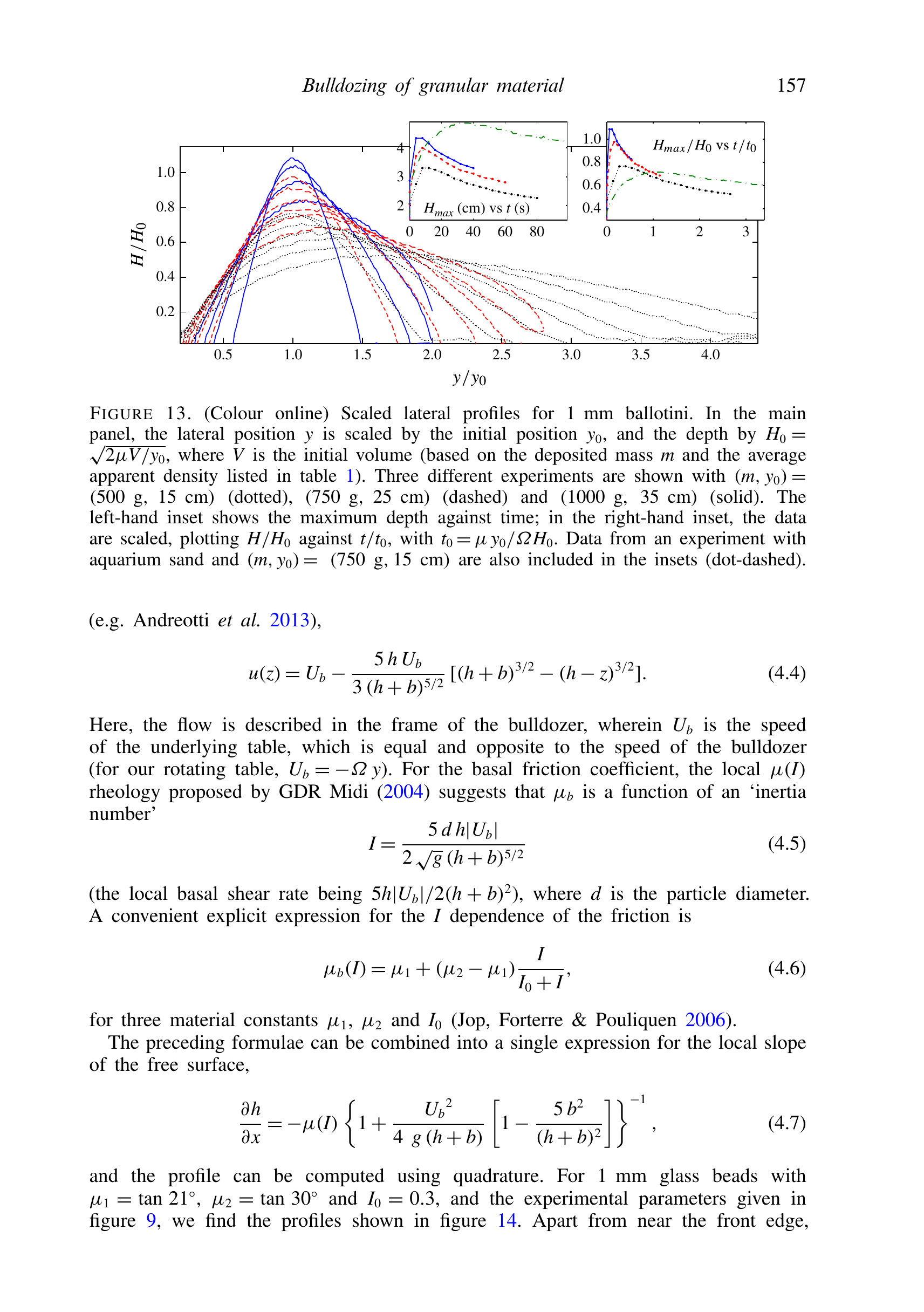}
    \caption{ 
Scaled lateral profiles for 1 mm ballotini. In the main panel, 
lateral position $y$ is scaled by the initial position $y_0$, and depth
by $H_0=\sqrt{2\mu V/y_0}$, where $V$ is the initial volume (based on the
deposited mass $m$ and the average apparent density listed in
table \ref{tableau_1}). Three different experiments are shown with
$(m,y_0)=$($500\,$g, $15\,$cm) (dotted), ($750\,$g, $25\,$cm) (dashed) and ($1000\,$g, $35\,$cm) 
(solid). The left-hand inset shows the maximum depth against time;
in the right-hand inset, the data is scaled, plotting
$H/H_0$ against $t/t_0$, with $t_0=\mu \,y_0 / \Omega H_0$.
Data from an experiment with aquarium sand and $(m,y_0)=$($750\,$g, $15\,$cm)
is also included in the insets (dot-dashed).
}
    \label{fig:temp1mm}
  \end{center}
\end{figure}

Lateral spreading also appears to be insensitive to the 
initial shape of the dune, as we found in other experiments in which
grains of aquarium sand were deposited in half-cones propped up against
the blade before rotating the table; once rotation began, the  
profiles spread laterally and quickly became very similar to those 
recorded for the full sandpiles that suffered collisions against the blade.
This insensitivity suggests a scaling
of the lateral profiles based on the depth-averaged model of
\S 4.2. According to this model, if
we scale the lateral position by $y_0$ and the profile by
a characteristic depth $H_0=\sqrt{2\mu V/y_0}$, 
then experiments with comparable
values of the parameter $C=\mu\delta/H_0$ should be similar at
the scaled time $t/t_0$, with $t_0=\mu y_0 / \Omega H_0$.
Here, $V$ is the
initial volume (estimated as $m/\rho$, with $\rho$ the average of the
loose and compacted apparent densities listed in table \ref{tableau_1})
and $\mu\equiv\mu_M$ is the leading-order perpendicular
slope.
The scaling is illustrated in figure \ref{fig:temp1mm} for three
of the $1$ mm ballotini experiments with $C\approx0.2$). 
The comparison is marred to begin with 
by the transient during which the 
memory of the detailed initial shape is destroyed. Beyond that transient,
however, the scaled profiles do appear to follow a common evolution
(at least until the dune is swept out to the edge of the table).
Figure \ref{fig:temp1mm} also includes data from one of the 
aquarium sand experiments (again with $C\approx0.2$); 
although the comparison
of the scaled data is less satisfying than for the ballotini,
the scaling enjoys some success in view
of the relatively large differences in the unscaled data.

\section{Theoretical models}

To parallel our experimental discussion, we now provide two versions
of a simple depth-averaged model designed to examine quasi-steady
perpendicular avalanching and lateral spreading.  We then present 
DEM simulations of a slightly simpler arrangement, the
planar bulldozer (\textit{i.e.} granular dunes driven forwards by a
vertical plate undergoing rectilinear motion), addressing some of the
limitations of the depth-averaged models.

\subsection{A depth-averaged model for the perpendicular
  profile}\label{sec:transversemodel}

\def\UU{{U}} \def\Ub{{U_b}}

Depth-averaged models are a conventional tool to analyse shallow,
incompressible granular flows (Savage \& Hutter 1991; Forterre \&
Pouliquen 2008; Andreotti, Forterre \& Pouliquen 2013).  In two
dimensions and for steady flow over a horizontal plane, the depth-integrated
equations expressing conservation of mass and horizontal momentum are
\begin{equation}
  \frac{\partial}{\partial x}\left(\int_{-b}^h u \,\,\text{d}z\right) = 0
\qquad {\rm 
and
} \qquad
  \rho\,\frac{\partial}{\partial x}\left(\int_{-b}^h u^2 \,\,\text{d}z\right) 
= 
 \frac{\partial}{\partial x}\left(\int_{-b}^h \sigma_{xx} \,\,\text{d}z\right) 
  - \mu_b\, p(x,-b) ,
  \label{B.2}
\end{equation}
where $u(x,z)$ denotes the horizontal velocity, 
the underlying plane is positioned at
$z=-b$ (the base of the blade of the bulldozer being located at $z=0$)
and $\rho$ is the density. 
Here, the normal stresses are $(\sigma_{xx},\sigma_{zz})$
and the shear stress on the base of the layer is expressed
as a product of the isotropic pressure $p(x,z)$ and a friction
coefficient $\mu_b$. For shallow flow, hydrostatic balance typically
prevails in the vertical direction and the pressure dominates the normal stresses,
\begin{equation}
\sigma_{xx}\approx\sigma_{zz}\approx-p
  \approx -\rho g (h-z), \label{eq:pdef}
 \end{equation}
where $g$ is gravity and we ignore the atmospheric pressure above the
granular layer. 

The first relation in (\ref{B.2}) indicates that the horizontal
flux is uniform, which, in view of the upstream conditions,
implies 
 \begin{equation}
\int_{-b}^h u \,\,\text{d}z = b\, \Ub.
 \end{equation}
To deal with the second relation, we require expressions for the
local vertical profile of the velocity and $\mu_b$.  For the former
we adopt the Bagnold form \cite[{\it e.g.}][]{bookpouliquen},
 \begin{equation}
   u(z) = \Ub-\frac{5\,h\,\Ub}{3\,(h+b)^{5/2}} \,[ (h+b)^{3/2} - (h-z)^{3/2} ] 
   .
   \label{B.5}
 \end{equation}
Here, the flow is described in the frame of the bulldozer, wherein $\Ub$
is the speed of the underlying table, which is equal and opposite to
the speed of the bulldozer (for our rotating table, $U_b=-\Omega\,y$).
For the basal friction coefficient, the local $\mu(I)$ rheology
proposed by \cite{gdr2004} suggests that $\mu_b$ is a function of an
 ``inertia number''
\begin{equation}
   I = \frac{5\,d\,h|\Ub|}{2\,\sqrt{g}\,(h+b)^{5/2}} 
   \label{B.8}
 \end{equation}
(the local basal shear rate being $5h|\Ub|/2(h+b)^2$), where $d$ is the
particle diameter.  A convenient explicit expression for the
$I$ dependence of the friction is
 \begin{equation}
   \mu_b(I) = \mu_1 + (\mu_2-\mu_1)\frac{I}{I_0+I} ,
   \label{B.9}
 \end{equation}
for three material constants $\mu_1$, $\mu_2$ and $I_0$ \cite[][]{jop2006}.

\begin{figure}
  \begin{center}
    \includegraphics[width=13cm]{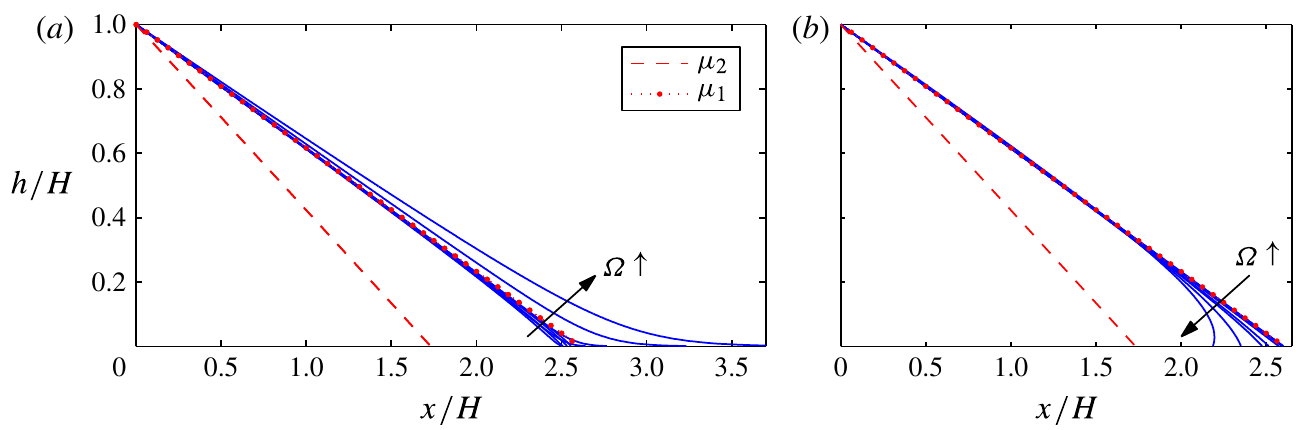}
    \caption{Perpendicular profiles computed from (\ref{B.10}) assuming
      $\mu_1=\tan 21^\circ$, $\mu_2=\tan 30^\circ$, $I_0=0.3$, $k=1$,
      $h(0)=H=5$\,cm, $r_0=25$\,cm and the same rotation rates as in
      figure~\ref{fig:increase_F}, i.e. $\Omega=0.05, 0.1, 0.2, 0.4,
      0.8, 1, 1.5$ and $2$\,rad\,s$^{-1}$. In (a) $b=0$; in (b) $b=2$
     \,cm.  }
    \label{fig:profile_perp_F_0}
  \end{center}
\end{figure}

The preceding formulae can be combined into a single expression for
the local slope of the free surface,
\begin{equation}
  \frac{\partial h}{\partial x} = - \mu(I) \left\{
    1 + \frac{{U_b}^2}{4\,g\,(h+b)} \left[1-\frac{5\,b^2}{(h+b)^2}\right]
  \right\}^{-1},
  \label{B.10}
\end{equation}
and the profile can be computed using quadrature.  For 1 mm glass beads with
$\mu_1=\tan 21^\circ$, $\mu_2=\tan 30^\circ$ and $I_0=0.3$, and the
experimental parameters given in figure~\ref{fig:increase_F}, we find
the profiles shown in figure~\ref{fig:profile_perp_F_0}. Apart near
the front edge, the free surface has a slope close to $\mu_1$.
Closer to the front, the shape depends on the depth of the incoming
layer, $b$, and the local Froude number $U_b/\sqrt{g(h+b)}$.  For
bulldozing on a ``dry'' plane ($b=0$) and relatively slow flows with
$U_b/\sqrt{g\,h}\ll1$, the front steepens up to a ``contact angle'',
$\tan^{-1}\mu_2$ \cite[{\textit{cf.}}][]{pouliquen1999b}. With inertia however, i.e. $U_b/\sqrt{g\,h}=O(1)$, material is pushed out ahead of the main
wedge, creating a distinctive forward skirt.  The situation is quite
different when there is an incoming layer, $b>0$: the factor
$1-5\,b^2/(h+b)^2$ reverses the sign of the inertial correction near the
front, with the result that the wedge steepens further there, and even
becomes vertical if the bulldozing speed is increased beyond
$\sqrt{g\,b}$, suggesting that the front overturns and breaks.

Although the depth-averaged model predicts that the profile is nearly
linear with a front that steepens with an increase in rotation rate,
much as in the experiments, the finer details of the measured profiles
are not reproduced (such as the flattening of the profile at
the blade). In fact, it is surprising that the model looks to
work better for the sand than the ballotini, despite the fact that the
$\mu$($I$) model is designed for the latter and is known to require
revision for sand \cite[{\it e.g.}][]{forterre2008}.  Part of
the problem is that the depth-averaged model discards horizontal
derivatives, with the consequence that there are sudden switches in
the flow pattern at the junction of the wedge with the incoming layer
and at the bulldozing plate. More properly, boundary layers should be
introduced at these locations to provide smoother connections. As is
clear from the DEM simulations described later, the assumption that
the vertical flow profile takes the Bagnold form in (\ref{B.5})
can also be suspect,
with the flow adjusting more gradually from the incoming plug to the
recirculating wedge. Moreover, the DEM simulations reveal 
interesting variations in the packing fraction, which
may impact the local rheology, whereas the model is incompressible.

\subsection{Lateral spreading}\label{sec:lateralmodel}

To construct a simple model of lateral spreading,
we couple the depth-averaged mass conservation equation with some
approximations suggested by the observation that flow becomes
quasi-steady perpendicular to the blade. For simplicity, 
we also ignore any motion in the uniform
layer flowing underneath the blade ({\it i.e.} in $z<0$). Conservation
of mass then implies
\begin{equation}
  \frac{\partial h}{\partial t}+
  \frac{\partial}{\partial x}\left(\int_{0}^h u\,\text{d}z\right)+
  \frac{\partial}{\partial y}\left(\int_{0}^h v\,\text{d}z\right)=0,
\end{equation}
where $\left(u(x,y,z,t),v(x,y,z,t)\right)$ denotes the horizontal velocity field.
We rewrite this equation as
\begin{equation}
  \frac{\partial h}{\partial t}+\frac{\partial}{\partial x}(h\,\Ub+F_x)
  +\frac{\partial}{\partial y}(h\,\Vb+F_y)=0,
  \label{eq:conservation_2D_lateral}
\end{equation}
where $\left(\Ub,\Vb\right)$ is the velocity of the underlying layer
in the reference frame of the blade and
\begin{equation}
  (F_x,F_y)\equiv\int_0^h (u-\Ub,v-\Vb)\, \text{d}z
\end{equation}
denotes the flux due to the avalanching internal motion of the
granular material.

\def\XX{{X}}

As discussed above, the granular flow in the transverse $x$-direction
adjusts relatively quickly and the bulldozed dune becomes
quasi-steady.  Therefore, the net transverse flux in
(\ref{eq:conservation_2D_lateral}), namely $h\,\Ub+F_x$, must become small,
\begin{equation}
  F_x \approx - h\,\Ub. \label{eq:fub}
\end{equation}
Any residual transverse flux balances the slow time variation of
$h$ and the weaker lateral flux along the blade. Nevertheless, that
residual flux must vanish exactly at both the blade and the leading
front $x=X(y,t)$ of the dune, and so
$[h\,\Ub+F_x]_{x=0}=[F_x]_{x=\XX}=0$. Hence, to model lateral
spreading, whilst avoiding the need to construct the residual
transverse flux, we integrate (\ref{eq:conservation_2D_lateral}) over
the $x$-direction to obtain the relation
\begin{equation}
  \frac{\partial }{\partial t}\left(\int_0^{\XX}\,h\,\,\text{d}x\right)
  +\frac{\partial }{\partial y} \int_0^{\XX}\,(h\,\Vb+F_y)\,\,\text{d}x
  = 0.
  \label{eq:spread_x}
\end{equation}

We now exploit the fact that the transverse profile of the dune is
almost linear,
\begin{equation} \label{eq:cond_slope} h(x,y,t)\approx H(y,t)-\mu\,x ,
\end{equation}
with constant slope, $\mu$. Hence $X(y,t)=H(y,t)/\mu$.  Furthermore,
for a free-surface gravity-driven flow, one would expect that the avalanche
flux, $(F_x,F_y)$, would be directed downslope, i.e.  $\mathbf{F} \approx -
\Gamma \nabla h$, where the factor $\Gamma$ encapsulates the detailed
physics of the granular flow.  Hence,
\begin{equation}
  F_y \approx F_x \,\frac{h_y}{h_x} \approx \frac{h\,\Ub\,H_y}{\mu}
\end{equation}
using (\ref{eq:fub}) and (\ref{eq:cond_slope}). Thus,
\begin{equation}
  \frac{1}{2\mu} \frac{\partial H^2}{\partial t}
  +\frac{\partial }{\partial y} \int_0^{{\XX}}\,
  (H-\mu x)\left(\Vb+\frac{\Ub\,H_y}{\mu}\right)\,\,\text{d}x
  = 0.\label{eq:latspreadbase}
\end{equation}
Once the basal velocity, $(\Ub,\Vb)$, is prescribed, the integral in
(\ref{eq:latspreadbase}) can be computed and the model can be completed.

\subsubsection{The planar bulldozer}

\begin{figure}
  \begin{center}
    \includegraphics[width=0.9\textwidth]{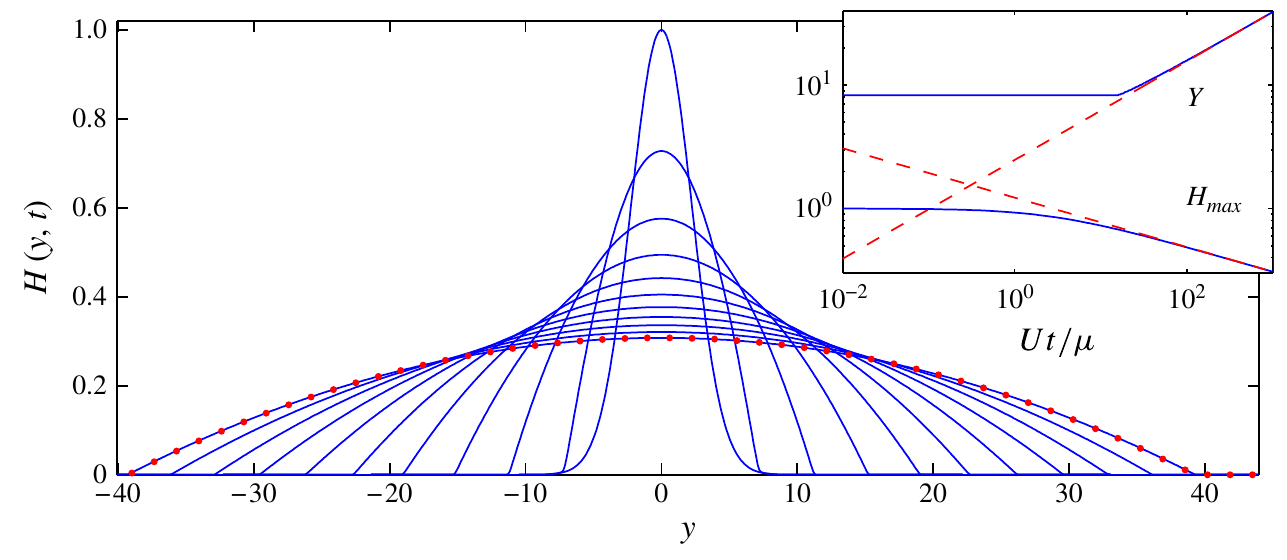}
     \caption{ Solution to the planar bulldozer problem defined by
      (\ref{pb}), showing the collapse of a Gaussian,
      $H(y,0)=e^{-y^2/10}+10^{-4}$, towards the similarity solution in
      (\ref{eq:pbsim1})-(\ref{eq:pbsim2}) (red dots and dashed lines).
      The main panels shows ten snapshots of the solution, equally
      spaced in $t^2$ up to $t=10^3\mu/|\Ub|$; the inset shows the
      maximum depth $H_{max}(t)$ and edge position $Y(t)$.  The
      pre-wetted film of depth $10^{-4}$ is added to avoid any
      numerical problems with the vanishing of the diffusivity in
      (\ref{pb}), and is sufficiently shallow that it has no effect on
      the solution.  For the numerical solution, the edges of the
      dune, $y=\pm Y(t)$, are defined as the positions where
      $H(y,t)=2\times10^{-4}$.\label{fig1}}
  \end{center}
\end{figure}

When the bulldozer undergoes rectilinear motion in the $x-$direction,
$\Vb=0$ and $U_b$ is a constant that must be negative if the blade is 
located at $x=0$ and the wedge is piled up in $x>0$. 
Equation (\ref{eq:latspreadbase}) then reduces to
\begin{equation}
  \label{pb}
  \left(H^2\right)_t = -\frac{\Ub}{\mu} \left(H^2 H_y\right)_y  .
\end{equation}
As illustrated in figure~\ref{fig1}, we may solve
(\ref{pb}) as an initial-value problem given a
suitable initial condition (a Gaussian in the figure). Over longer
time, the numerical solution converges to a similarity solution given
by
\begin{equation}
  H(y,t)=\left\{ \begin{array}{lc}
      H_{max}(1-y^2/Y^2), & {\rm for}\ -Y < y < Y,\cr
      0 , & {\rm elsewhere,}
    \end{array}\right.
  \label{eq:pbsim1}
\end{equation}
where
\begin{equation}
  Y(t) = \left(\frac{375\,V \,\Ub^2\, t^2 }{8\,\mu}\right)^{1/5},
  \qquad
  H_{max}(t)= \left(\frac{45\,\mu^3 \,V^2} {64 \,t\; |\Ub|}\right)^{1/5} . 
  \label{eq:pbsim2}
\end{equation}
and $V\equiv \int H^2 \text{d}y / 2\mu$ is the volume of grains in the
dune. Solutions beginning with a wide range of single-humped initial
conditions converge to the self-similar form in (\ref{eq:pbsim1}).  In
other words, after a transient that obliterates the initial shape, the
mound adopts a characteristic parabolic profile, spreading laterally
like $t^{2/5}$ {while} its maximum depth {falls} like $t^{-1/5}$.

\subsubsection{The rotating bulldozer}

In our Cartesian coordinate system, the front face of the blade lies
along the $(y,z)-$plane at $x=0$. However, as already noted, in our
experiments the rotation axis was offset from the origin of the
coordinate system by a distance $\delta$. The basal velocity field is
then
\begin{subeqnarray}
  \Ub & = & -\Omega\,y, \\
  \Vb & = & \Omega\,(x+\delta)
\end{subeqnarray}
(the rotation axis being located along the line
$(x,y,z)=(-\delta,0,z)$).  Thus,
\begin{equation}
  \frac{\mu}{\Omega}
  \frac{\partial H^2}{\partial t}+
  \mu\,\delta \,\frac{\partial H^2}{\partial y} 
  = \frac{1}{3}\left[ y^2 \left(\frac{H^3}{y}\right)_y\right]_y .
  \label{eq:advectiondiffusion}
\end{equation}
It should be noted that (\ref{eq:advectiondiffusion}) applies only for $y>0$; for
$y<0$, the granular medium must be piled up on the opposite side of
the blade, which calls for some key switches of sign in the formulae.

\begin{figure}
  \centering
  \includegraphics[height=5.5cm]{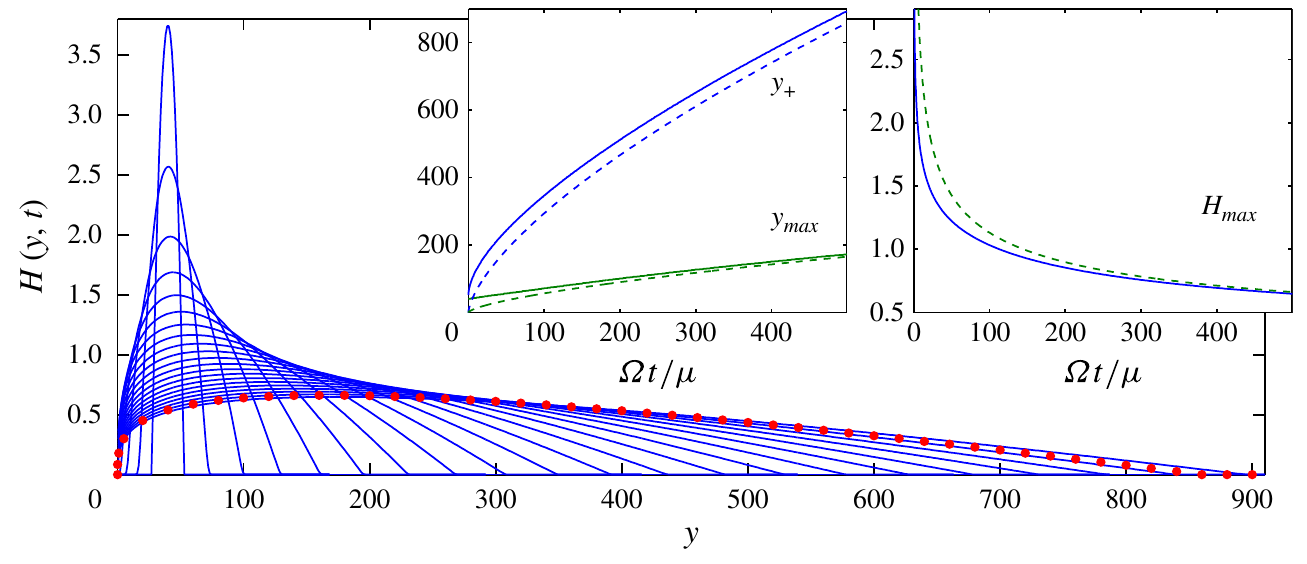}
  \caption{ Numerical solution to the rotating bulldozer problem in
    (\ref{eq:advectiondiffusion}) without radial advection
    ($\delta=0$), showing the collapse of a parabolic mound (again
    supplemented by a thin pre-wetted film of depth $10^{-4}$) towards
    the similarity solution in (\ref{eq:ssrot1})-(\ref{eq:ssrot2})
    (red dots and dashed lines).  The main panels shows twenty
    snapshots of the solution, equally spaced in $t^2$ up to
    $t=500\mu/\Omega$; the insets show the position of the 
    right-hand edge of the dune,
    $y_+$, the location of the maximum depth, $y_{max}$, and
    that depth, $H_{max}(t)$.  For the numerical solution, $y_+$
    is defined by $H(y_+,t)=2\times10^{-4}$.}
  \label{fig2}
\end{figure}

We first consider the case when the blade is positioned exactly along
a diameter of the rotating table, so that $\delta=0$.
Numerical solutions to the corresponding initial-value problem
illustrate how the dune slumps preferentially radially outwards; see
figure~\ref{fig2}, which shows the lateral spreading of an initially
parabolic mound.  The granular material also piles up towards the
centre, building up a sharp edge near $y=0$. Again the solution
converges to a self-similar solution, this time given by
\begin{equation}
  H = \left\{ \begin{array}{lc}
      \mu \, t^{-1}\, \Omega^{-1} \, y^{1/3}  \, ( Y^{2/3} - y^{2/3} ),  & 0<y<Y, \cr
      0, & {\rm elsewhere},
    \end{array}\right.
  \label{eq:ssrot1}
\end{equation}
where
\begin{equation}
  Y=\left( \frac{105 \, t^2 \, V \, \Omega^2}{ 4 \, \mu}\right)^{1/3}
  \label{eq:ssrot2}
\end{equation}
denotes the lateral extent of the dune, in terms of which the maximum
depth and its location are given by
\begin{equation}
  H_{max}= \frac{2\, \mu Y}{3\sqrt3\;t\,\Omega}
  \qquad {\rm and} \qquad
  y_{max}=\frac{Y}{3\sqrt{3}} .
\end{equation}
Evidently, the linear rise of the rotation speed with $y$ increases
the rate at which the dune is swept outwards and thins ($Y\sim
t^{2/3}$ and $H_{max}\sim t^{-1/3}$ rather than $t^{2/5}$ and
$t^{-1/5}$, respectively, for the planar bulldozer).

\begin{figure}
  \begin{center}
    \includegraphics[width=\textwidth]{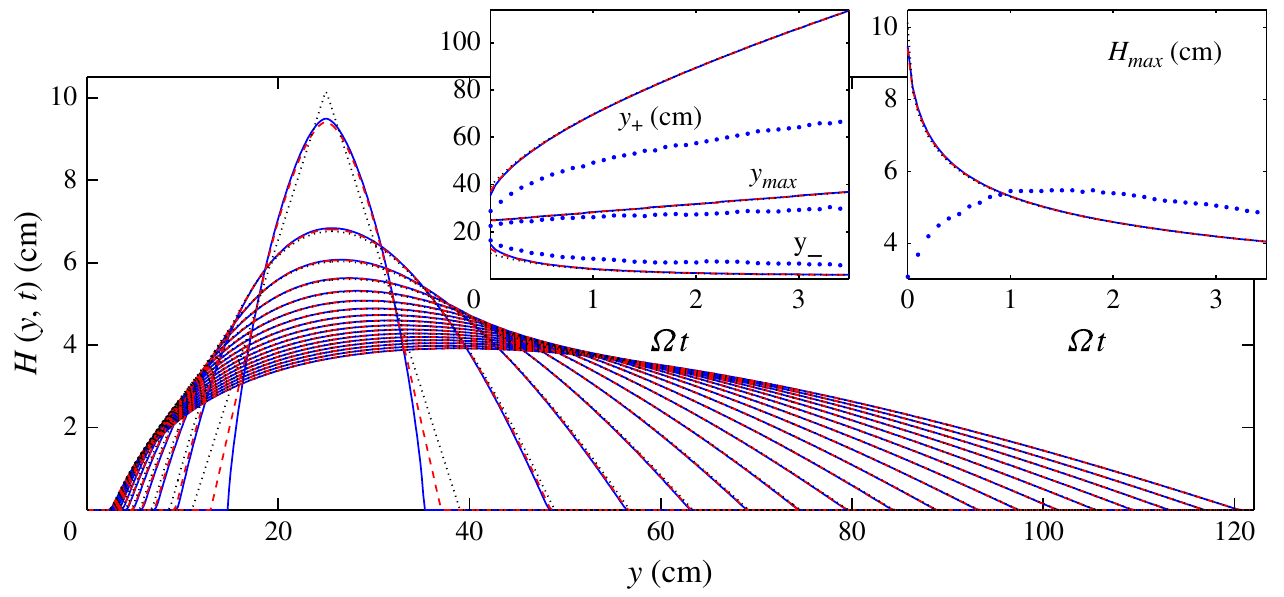}
    \caption{Numerical solution to (\ref{eq:advectiondiffusion}) with
      $\Omega=0.05$\,rad\,s$^{-1}$, $y_0=25$\,cm, $\mu=\tan 36^\circ$
      and $\delta= 2$\,cm. The snapshots of $H(y,t)$ are shown every 5
      seconds, and the insets show the positions of the dune's maximum
      $y_{max}(t)$, the edges of the dune $y_\pm(t)$, and the maximum
      depth $H_{max}(t)$. Three solutions are shown, starting from three different initial conditions but all with the same mass: a triangle (dotted), the profile expected from rearranging a sandpile (solid) and the profile obtained when the rearranged sandpile is adjusted to limit its maximum slope (dashed), as described further in the main text.
  In each case, a pre-wetted film of depth $10^{-4}$ is added and the edges are defined as
      the locations where $h(y,t)=2\times10^{-4}$.  The dots in
the insets show the corresponding data from the experiment in
figure \ref{fig:profile_syst_mass_what_211}b.
}
    \label{Num_Theo_profile}
  \end{center}
\end{figure}

When $\delta\ne0$, the flux along the blade picks up an additional
component that helps to sweep the dune out to larger radii, where
rotational bulldozing effects a rapid diffusion to flatten out the
mound. The additional advection prevents material from piling up
towards the centre of the rotating table, and the solution no longer
converges to the similarity solution (\ref{eq:ssrot1}) and
(\ref{eq:ssrot2}).  The dynamics is illustrated by the solutions of
the initial-value problem shown in figure \ref{Num_Theo_profile}.
These examples use parameter settings based on the experiment with
aquarium sand shown in figure \ref{Num_Theo_profile} ($\delta= 2$\,cm
and $\mu =\tan\theta_M\approx \tan 36^\circ$), and use three different
initial profiles $H(y,0)$, all with the same volume. The first is a
profile obtained by taking a sandpile with slope
$\mu_s\approx\tan33^\circ$ centred at $y_0$ and then rearranging the
material at each $y$ into wedges with slope $\mu$; this is the
theoretical initial condition corresponding to the experiment given
our assumption of a relatively rapid adjustment to a quasi-steady
state of perpendicular avalanching. The second initial profile is also
based on rearranging the sandpile. However, the simplest rearrangement
of this sort
furnishes an initial shape with slopes that exceed $\mu$, which would
presumably avalanche laterally; to avoid this detail, the steeper
slopes of the rearranged sandpile are replaced with linear sections
of slope $\mu$, and the entire profile then rescaled to furnish the
correct volume. The third initial condition is simply a triangle with
slopes of $\mu$.  As can be seen in the figure, after a short
transient (lasting less than 10 seconds), all three cases converge to
the same spreading solution.  Thus, the finer details of the initial
profile are not significant, as found experimentally.

Qualitatively, the solutions in figure~\ref{Num_Theo_profile}
are similar to the experimental observations.  
However, although the theory rationalizes the collapse of the
experimental data in figure~\ref{Num_Theo_profile_b}
for different rotation rates, 
the agreement between the model and the experiment is not quantitative.
This is shown in the insets of figure~\ref{Num_Theo_profile},
which compare the theoretical predictions for
the positions of the edges and maximum of the dune
and its maximum depth with data from the experiment in
figure \ref{fig:profile_syst_mass_what_211}b.

The insensitivity to initial shape suggests
a convenient rescaling of (\ref{eq:advectiondiffusion}). If we set 
\begin{equation}
\label{scal}
\hat{y} = \frac{y}{y_0},\qquad
\hat{H} = \frac{H}{H_0},\qquad
\hat{t} = \frac{t}{t_0} , \qquad
H_0=\sqrt{\frac{2\mu V}{y_0}} 
\qquad {\rm and} \qquad
t_0 = \frac{\mu \, y_0}{\Omega \, H_0} ,
\end{equation}
then the evolution equation becomes
\begin{equation}
    \frac{\partial \hat{H}^2}{\partial \hat{t}} +
  C \, \frac{\partial \hat{H}^2}{\partial \hat{y}} 
  = \frac{1}{3}\left[ \hat{y}^2 \left(\frac{\hat H^3}{\hat y}\right)_{\hat y}
\right]_{\hat y} 
\end{equation}
and the rescaled volume constraint is $\int \hat H^2 {\rm d}\hat y = 1$.
Thus, the later-time solutions depend on the single parameter $C$,
as was roughly confirmed experimentally in \S 3.3.

\subsection{DEM simulations}

For our DEM simulations, we used identical deformable, spherical
particles. All variables are expressed in dimensionless form, with
lengths scaled by the particle radius $d/2$, speeds by $\sqrt{g\,d/2}$ and
times by $\sqrt{d/2g}$.  Particle collisions are dealt with using a
soft-particle model with a damped linear spring for the normal force
(with zero coefficient of restitution) and dynamic Coulomb friction
for the tangential force (with a coefficient of friction equal to
0.3). The particle
stiffness was chosen so that any overlap would not exceed 1\%.  The
time step was one tenth of the binary collision time; additional
computations with a smaller time step verified the insensitivity of the
results to this choice.
Simulations were also performed with different stiffness values and with
a static friction model wherein the
tangential displacement of each contact was tracked; there was no 
discernible change in the main characteristics of the dunes.

To mimic the sandpaper glued to the
surfaces of the experimental bulldozer and the rotating table, we suitably
positioned a set of immobile particles to create a rigid vertical
plate and translating underlying plane. To help prevent the particles from crystallizing
against these boundaries, the immobile particles were placed on a
rectangular grid with a spacing equal to $\Delta=3(2+\sqrt 3)/5$ times
the particle radius. We performed two suites of computations: in one,
the base of the bulldozing plate was located
20 particle radii above the plane; for the other, 
the base of the blade was placed next to the plane to avoid
any gap between the two obstructions. In the latter,
particles do become temporarily jammed between the bulldozer and the surface
(the effect
we sought to avoid in the experiments by emplacing the underlying layer);
when released, these grains gain significant kinetic energy
and can become launched out on ballistic trajectories.
This dynamics depends on the model for particle collision 
but does not affect the overall behaviour of the dunes.

Unlike the rotational geometry of the experiment, the DEM
simulations
were conducted with a bulldozing plate undergoing rectilinear perpendicular 
motion in Cartesian geometry. We used two configurations. First,
to explore the perpendicular dynamics independently of any lateral
spreading, we used a relatively narrow slot 
with a perpendicular length of $150\Delta$ and a
lateral width of $5\Delta$ (corresponding to about 150 by 6
particles). The domain was taken to be periodic in both horizontal 
directions, and 20\,000 or 10\,000 particles were released,
depending on whether there was an underlying layer or not.
The narrowness of the slot eliminates coherent lateral motion, and
simulations in domains of different width suggested little dependence
on that dimension. In our {initial} simulations, the particles were
deposited in a triangular wedge; the ensuing bulldozing flow led to a
vigorous adjustment, with a fraction of the {particles} launched away
from the bulk on ballistic trajectories. The material took some time
to settle down after this transient (requiring several
passes of the blade through the domain to allow the wedge to adjust
and the packing in the bed to increase). In subsequent simulations, 
we therefore used the
particles positioned from an earlier evolved solution with a
different bulldozer speed.

Second, to study lateral spreading, we prepared a much
wider simulation by assembling several side-by-side copies of one of
the narrow simulations and evolved this configuration
for a while to reduce any spatial
correlations. We then formed a localized mound against the blade in
the centre of this domain by taking the profile along the midline
$y=0$ and rotating this curve about the origin, clipping all the
overlying particles. This created a half-cone above the bed (as can be
seen in the left panel in figure~\ref{fig:dem3d}) and minimized any
subsequent vigorous adjustment.  The width of the computational domain
(\textit{i.e.} the number of copies of the original slot) was chosen to
be sufficiently large compared with the half-cone so that 
mobilized particles did not reach
the lateral boundaries, thus rendering irrelevant the precise width
and the lateral boundary conditions.  
Altogether, there were about $1.2$ million particles in the 
lateral spreading simulations with an underlying layer and about
 180\,000 particles in the simulations without one.

\subsubsection{Dynamics of perpendicular avalanching}

In our dimensionless units and with a fixed number of particles, there
is only a single parameter in the simulations, which is the bulldozer
speed $|U_b|$. During bulldozing, the particles become distributed
so that they build up a wedge of depth $H$ against the blade, furnishing the
effective Froude number of the flow, $|U_b|/\sqrt{H}$. Our main
interest is in characterizing
the steady flow states, so we omit any discussion of the
initial transients (which can be seen in the movies in the
supplementary material). Nevertheless,
the nature and existence of a steady flow state is not straightforward.
At the lowest Froude number, the flow becomes more intermittent in time.
At higher speeds, flow patterns can also vary
over very long time periods (hundreds of passes of the bulldozer). 
These variations can be due to small increases in
the packing fraction in the basal layer and the dune, which can lead to 
flows with much larger fluctuations in the forces and velocities. This
is presumably the same effect as seen by \cite{gravish2010}
in experiments with over consolidated packings.
Moreover, despite our efforts to avoid {crystallization}, arrangements
of perfectly close-packed grains can form adjacent to the blade
and generate long-time-scale variability,
as described below. Nonetheless, these problems do not
substantially affect the results we present and do not
mar our lateral spreading simulations.

\begin{figure}
  \centering
  \includegraphics[width=0.9\textwidth]{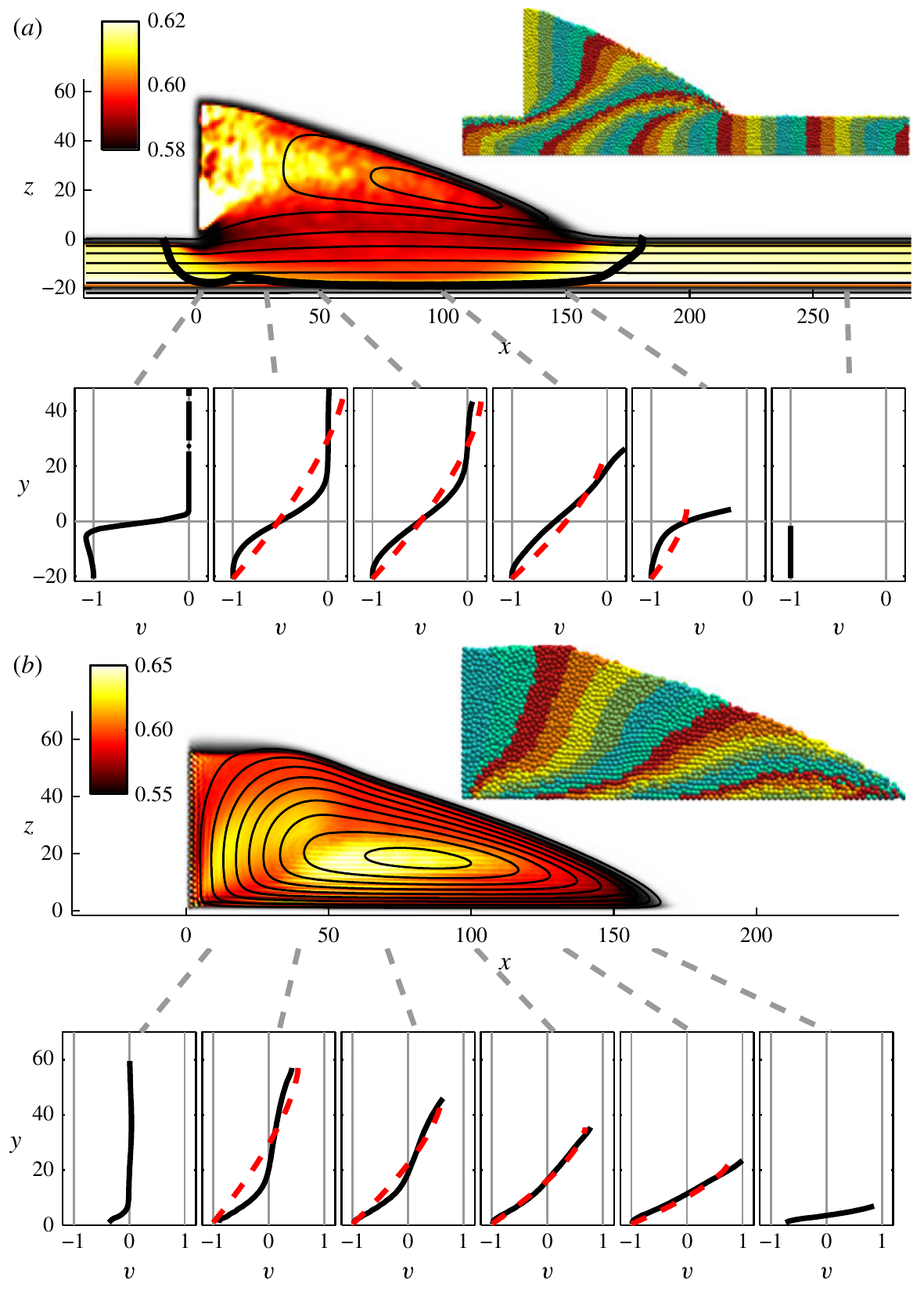}
  \caption{The steady-state flow pattern set up in simulations
    in the narrow slot for $|U_b|=1$ with an underlying layer in (a) and without one in (b).
    The black lines in the main panels show a selection of streamlines (not
    equally spaced); superposed is the packing 
    fraction, plotted as a density on the $(x,z)-$plane. At the horizontal
  positions $x$ indicated, the vertical profile of the horizontal velocity $u$
    is plotted in the sequence of panels underneath; the dashed lines show
    corresponding Bagnold profiles.
    The insets show snapshots of the particle
    distribution, with particles colour-coded according to their
    horizontal position $T$ time units earlier; in (a) T=100, whereas T=50 in (b). 
    (A movie of the simulation without an underlying layer
    is available as supplementary material). }
  \label{fig:dem-stream}
\end{figure}

The results of simulations at speed $|U_b|=1$ with and without the
underlying layer are displayed in figure~\ref{fig:dem-stream}. 
The steady-state flow patterns within the wedges built up against
the blade are shown; 
the superposed density plots display the packing fraction
and the insets show snapshots of the particle positions colour-coded
by their location at a given time earlier. Also indicated are
a selection of vertical profiles of the horizontal velocity.  In
figure~\ref{fig:dem-stream}(a), for the simulation with an underlying
layer, the wedge is about 30 particles high, which is a little more
than in the largest ballotini experiment.  In agreement with the
experiments, the wedge has almost constant slope, flattening out near
the blade and steepening near the flow front. The simulation
without an underlying layer in figure~\ref{fig:dem-stream}(b)
furnishes a wedge that is even flatter at the blade and steeper at its
front, illustrating how the underlying layer influences the shape of
the free surface. The recirculation cell within
the wedge is also very different. The vertical profiles of the horizontal velocity are fairly
well reproduced by the Bagnold profile in the case without an
underlying layer, but the comparison is poorer when the layer is
present.  For either case, the vertical shear is finite at the
surface, unlike that predicted by the Bagnold profile. One should note the weak
backflow underneath the blade in figure~\ref{fig:dem-stream}(a).

The ``yield surface'' that divides the deforming wedge from the
``plug'' flow in the layers in front and behind is also shown in figure
\ref{fig:dem-stream}(a). For this simulation, there is a weakly
deforming triangular region propped up against the bulldozer which can
be identified by an irregular pattern in the packing fraction. The
weak deformation is revealed by the undeformed vertical stripes in
the colour-coded particle positions in the inset and by the flat
section in the second and third profiles of the horizontal velocity.
This region is problematic in the DEM simulation as particles
sometimes crystallize there. The rigid rotation of these persistent ``frozen
crystals'' dramatically impacts the flow field and surface shape,
and generates long-time-scale variability. As the
crystallization is prompted by conducting a relatively
narrow simulation with identical particles, we omit a description of its 
dynamics in any more detail.

\begin{figure}
  \centering
  \includegraphics[width=0.9\textwidth]{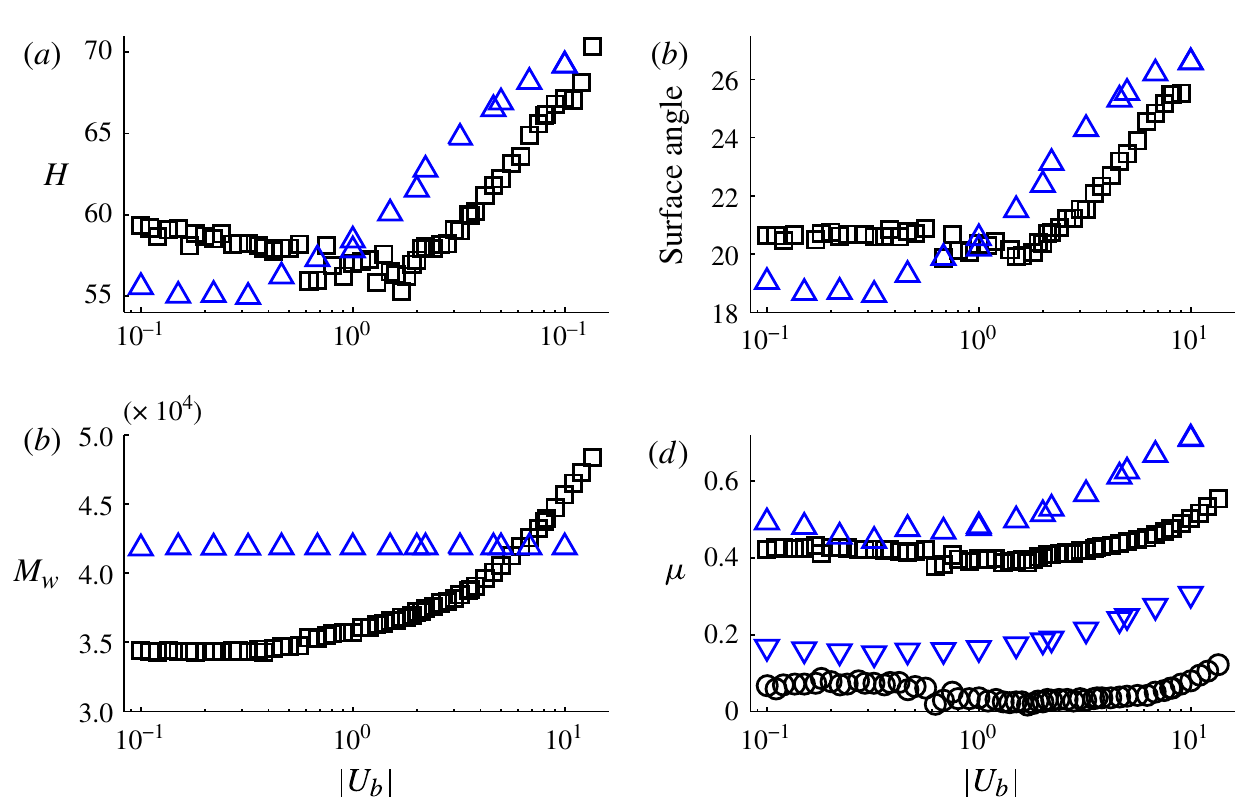}
  \caption{A summary of steady-state results for simulations in the
    narrow slot showing the dependence on $|U_b|$ of (a) depth at the
    blade, $x=0$, (b) the leading-order surface slope angle, (c) the
    mass of material in the wedge $M_w=\sum_j (u_j-U_b)/|U_b|$, 
and (d) the effective
    horizontal and vertical friction coefficients,
    $(\mu_x,\mu_z)=(F_x,F_y)/(gM_w)$, where $(F_x,F_z)$ is the net
    force on the blade ($F_x<F_z$).  The squares and circles show the results with
    an underlying layer; the triangles show the results without one.
  }
  \label{fig:dem-HqV}
\end{figure}

\begin{figure}
  \centering
  \includegraphics[width=0.8\textwidth]{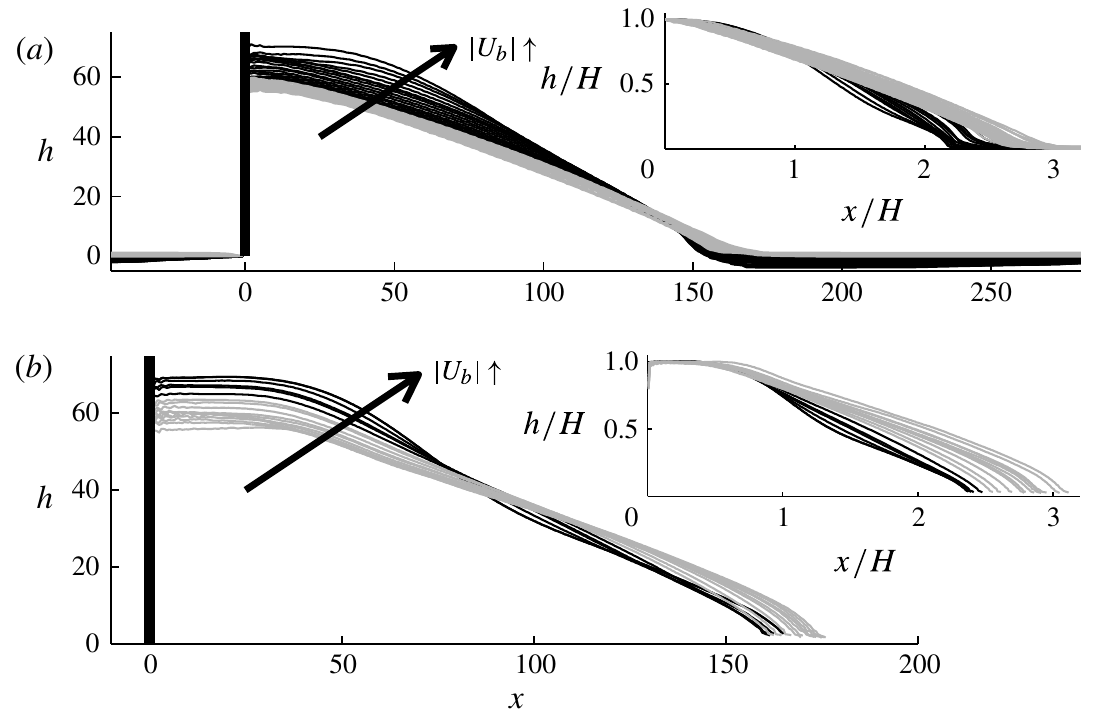}
  \caption{(a) Steady-state wedge profiles for simulations with
    varying bulldozer speed in the narrow slot (a) with 
and (b) without an underlying layer.
    The black lines denote the profiles for
    $|U_b|\geq 3.2$ (beyond the minimum of the curve in
    \ref{fig:dem-HqV}(a)); the grey lines denote the simulations
    with $|U_b|<3.0$. 
In the insets, the profiles are rescaled
    by the depth at the blade, $x=0$, to furnish the master curves,
after subtracting the level of the incoming layer in (a).
}
  \label{fig:dem-x-profile}
\end{figure}

Results from simulations with varying bulldozer speed are
summarized in figures~\ref{fig:dem-HqV} and \ref{fig:dem-x-profile}. 
Figure~\ref{fig:dem-HqV} plots how
various characteristics of the wedge depend on $|U_b|$;
figure~\ref{fig:dem-x-profile} shows
the corresponding wedge profiles.
With an underlying layer, the shape of the wedge is
largely independent of the Froude number for $\Fr<0.3$ ($|U_b|<3$);
see the data for the wedge height and slope in
figure~\ref{fig:dem-HqV}a,b, and the collapse of the profiles to a common
master curve shown in the inset of 
figure~\ref{fig:dem-profiles}a.
For $\Fr>0.3$ ($|U_b|<3$), the height, angle and profile of the wedge
depend noticeably on the Froude number; the Froude-number dependence of
the master curve at these bulldozer speeds is somewhat different from
that observed for the experiments. This discrepancy is partly
due to the lateral spreading of the experimental dunes 
({\it cf.} \S 4.3.2). However, the DEM simulations also have the feature
that the depth of the surface behind the bulldozer
decreases slightly as $|U_b|$ increases (see the left-hand edges of the
profiles in figure~\ref{fig:dem-profiles}a); faster wedges therefore
encounter shallower incoming layers, unlike in the experiments
(which last for one rotation of the table or less). The thinning
results from the backflow under the blade evident in
figure~\ref{fig:dem-stream}(a) and is described in more detail in
\cite{percier2011}. 
For the simulations without the underlying layer, there is a much
more gradual and monotonic dependence of the wedge characteristics
on the Froude number, and the wedge profiles do not collapse
as cleanly onto a common master curve for 
the range of Froude numbers simulated.

\subsubsection{Lateral spreading}

\begin{figure}
  \centering
  \includegraphics[width=\textwidth]{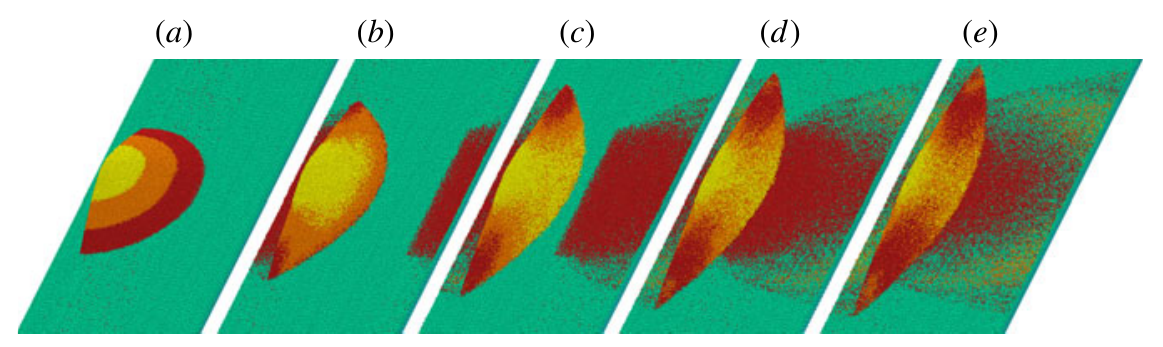}
  \caption{Particles positions at $t=0$, 100, 200, 300 and 400 for
a lateral spreading simulation with an underlying layer and
    $|U_b|=1$. The particles are coloured according to their initial
    height. (Video of this simulation is available as
      supplementary material).}
  \label{fig:dem3d}
\end{figure}

\begin{figure}
  \centering
  \includegraphics[width=0.9\textwidth]{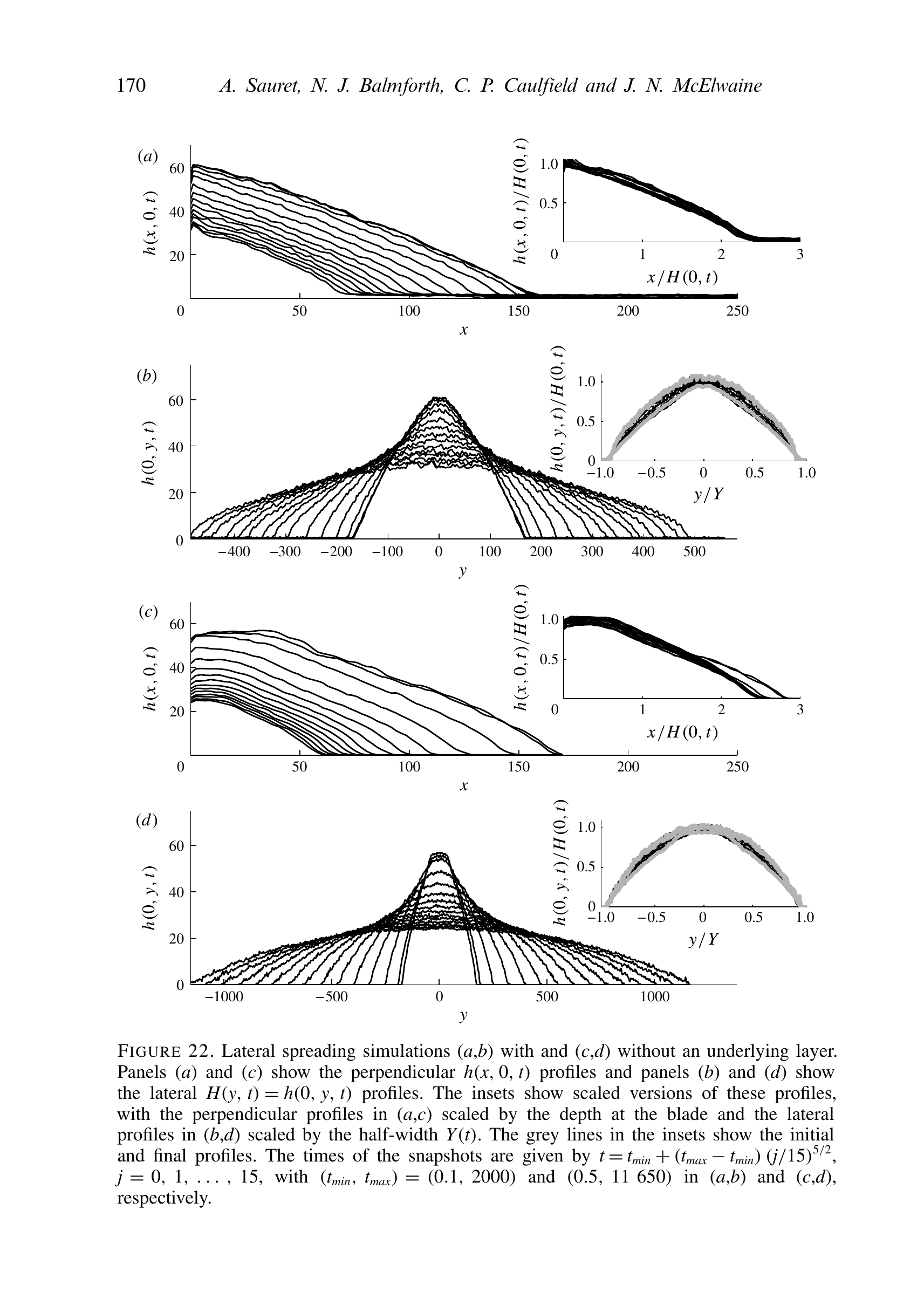}
  \caption{
Lateral spreading simulations (a,b) with and (c,d) without
an underlying layer. Shown are
(a,c) the perpendicular $h(x,0,t)$, and (b,d) the lateral
$H(y,t)=h(0,y,t)$ profiles.
The insets show scaled versions of these
profiles, with the perpendicular profiles in (a,c) scaled by depth at
the blade, and the lateral profile in (b,d) scaled 
by the half-width $Y(t)$.
    The grey lines in the insets show the initial and final profiles.
The times of the snapshots are given by
$t=t_{min}+(t_{max}-t_{min})\left({j}/{15}\right)^{5/2}$, $j=0$, 1, $...$, 15,
with $(t_{min},t_{max})=(0.1,2000)$ and $(0.5,11650)$ in (a,b) and (c,d), 
respectively.
}
  \label{fig:dem-profiles}
\end{figure}

The laterally spreading solution with an underlying plane
for $|U_b|=1$ is displayed in {figure
\ref{fig:dem3d}}. Because the bulldozing is planar in the simulation,
the spreading is symmetrical, unlike in the experiments. Despite
this difference, there are still 
qualitative similarities between the {simulations and experiments}.
Further details of the simulation are shown in
figure~\ref{fig:dem-profiles}, which {plots} snapshots of the
perpendicular profiles along the midsection of the dune (\textit{i.e.}
$y=0$) and its lateral profile along the blade (the depth at $x=0$) at
different times. These profiles appear to be self-similar, as
indicated by the insets of the figure, which scale the perpendicular
profiles with $H(0,t)$ and the lateral profiles with $Y(t)$, the
lateral extent of the dune.  Figure~\ref{fig:dem-profiles} also shows the
corresponding results for a simulation without an underlying layer.
Qualitatively, the character of the evolution is very similar. In particular, for the simulation without an underlying layer, the
self-similar lateral profile adopts a form that is well fitted by the
parabolic similarity solution in (\ref{eq:pbsim1}). It should be noted that the
perpendicular master profile shown in the inset of
figure~\ref{fig:dem-profiles}(a) is somewhat steeper at the blade
than the narrow-slot simulations in figure~\ref{fig:dem-x-profile}a, and
therefore compares more favourably with the experimental profiles.

\begin{figure}
  \centering
  \includegraphics[width=0.95\textwidth]{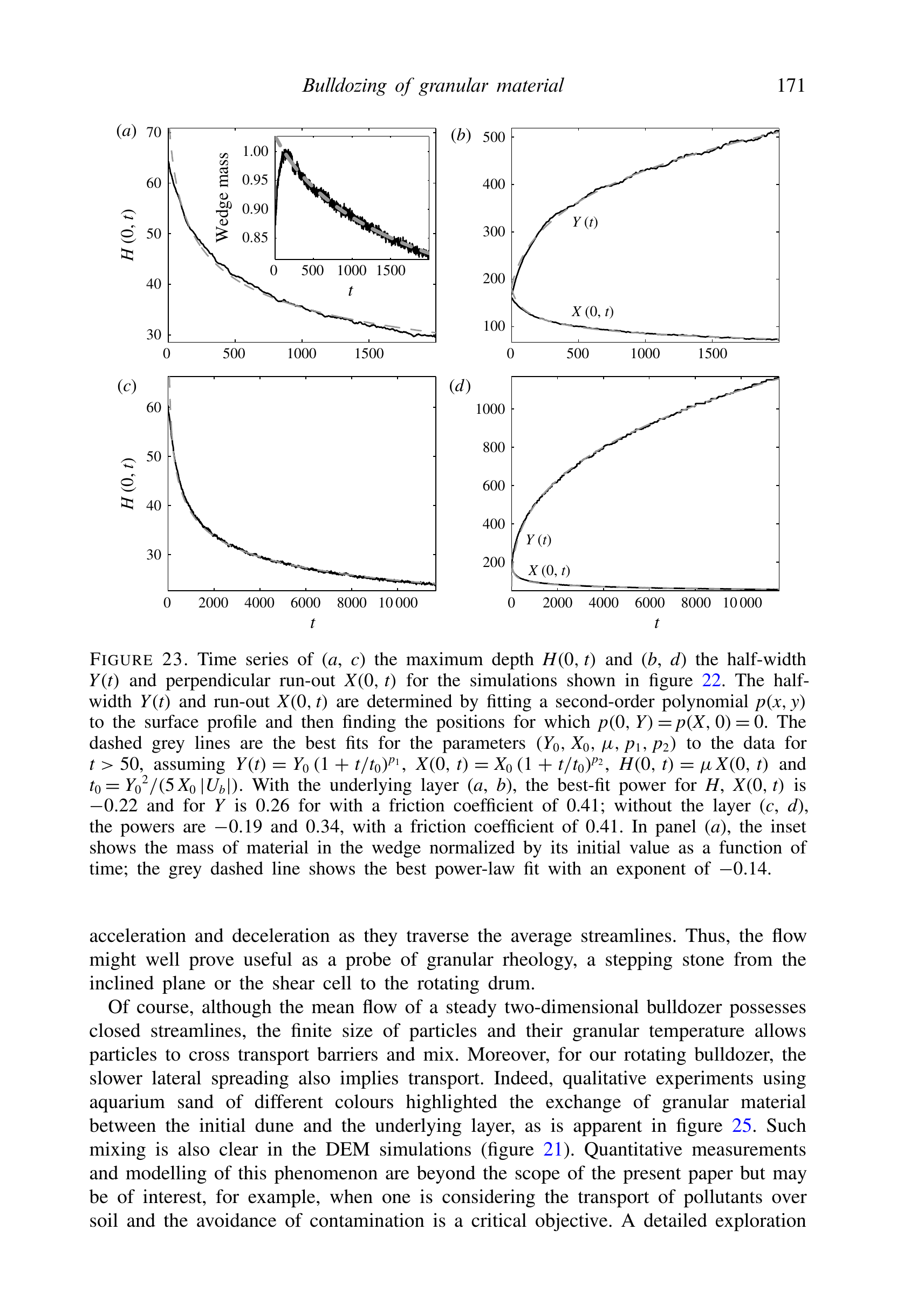}
  \caption{Time series of (a,c) maximum depth $H(0,t)$ and (b,d) half-width $Y(t)$ and perpendicular run-out $X(0,t)$, for the simulations shown in figure \ref{fig:dem-profiles}. The half-width $Y(t)$ and run-out $X(0,t)$ are determined by fitting a second order polynomial $p(x,y)$ to the surface profile and then finding the positions for which $p(0,Y)=p(X,0)=0$.  The dashed grey lines are the best fits for the parameters $(Y_0,X_0,\mu,p_1,p_2)$ to the data for $t>50$, assuming $Y(t)=Y_0\,(1+t/t_0)^{p_1}$, $X(0,t)=X_0\,(1+t/t_0)^{p_2}$, $H(0,t)=\mu\,X(0,t)$ and $t_0={Y_0}^2/(5\,X_0\,|U_b|)$ as predicted by the theory. With the underlying layer (a,b), the best fit power for $H$, $X(0,t)$ is $-0.22$ and for $Y$ is $0.26$ for with a friction coefficient of 0.41; without that layer (c,d), the powers are
$-0.19$, and $0.34$, with a friction coefficient of 0.41. In panel (a), the inset shows the mass of material in the wedge normalized by its initial value as a function of time; the grey dashed line shows the best power-law fit with an exponent $-0.14$.}
  \label{fig:dem-similarity}
\end{figure}

Figure~\ref{fig:dem-similarity} plots the time series of $H(0,t)$,
$Y(t)$ and the maximum perpendicular run-out, $X(0,t)$, for both
simulations in figure~\ref{fig:dem-profiles}. Also included are
best-fit power-law fits to these series. For $H(0,t)$ and $X(0,t)$,
the data are roughly fitted by $t^{-1/5}$, which is the self-similar
scaling predicted by (\ref{eq:pbsim2}b) (the best-fit powers are
$-0.22$ with the underlying layer and $-0.19$ without it).  However, although the best-fit power law for $Y(t)$ is
not too far from $t^{2/5}$ for the simulation without an underlying
layer (the best-fit power is $0.34$), it is quite different with the
basal layer (being $0.26$). The key to explaining this discrepancy is to
examine the total mass contained in the dune $M_w$:  for
the simulation without the underlying 
layer, the best-fit powers for $H$, $Y$ and
$X(0,t)$ sum closely to zero, confirming the notion that the dune is a
slowly spreading, but constant-mass structure.  With the underlying
layer, however, the exponents do not sum to zero, suggesting that the
dune is losing mass as time progresses. Indeed, direct measurements 
(figure~\ref{fig:dem-similarity}a) indicate a power-law decay of $M_w$
beyond an initial transient, with a best-fit exponent of $-0.14$
which is close to the summed
exponents of $H$, $Y$ and $X(0,t)$.  The mass loss
occurs because the backflow under the blade
effectively skims off material from the underlying layer and further
builds up the dune at the beginning of the simulation. Due to lateral
spreading, however, the backflow weakens with time, returning mass to
the underlying layer.  Evidently, the
scaling of $Y(t)$ is the main casualty, probably
because the shallow edges of the dune are affected most.

\section{Conclusion}
\label{conclusion}

In this paper, we have studied the granular flow created by bulldozing
a pile of grains over a level surface, both experimentally and
theoretically. In the experiments, we bulldozed a granular mound over
a rotating table, demonstrating how an avalanching dune is driven
forwards {in two phases}. A first rapid adjustment takes place
perpendicular to the blade, with the emergence of a quasi-steady
avalanching state characterized by a profile that has an
almost constant slope. {This} rapid adjustment is followed by a
second slower phase of lateral spreading parallel to the blade. {We}
constructed two simple depth-averaged {theoretical} models to describe
the perpendicular steady-state avalanching and lateral spreading.  The
first model captures the leading-order linear slope of the
experimental dunes, but not 
the finer details of their nonlinear shape. 
The experimental observations of asymmetrical radial spreading
and outward migration are matched qualitatively, but not quantitatively,
by the second model.

Our theoretical discussion is further shored up by DEM simulations,
which reproduce the phenomenology seen in the experiments and
offer a better comparison with some of the more detailed
observations. We have not, however, exploited the full power of the DEM
simulations in dissecting the flow dynamics non-invasively, or to
study the detailed effect of some of the physical parameters, 
such as the depth of
the underlying layer and the detailed boundary conditions on the
blade. Our computations do suggest that both of
these are key in determining the shape of the perpendicular profile at
the blade. As an alternative to DEM,
two-dimensional computations of the free-surface flow are also feasible
for fluid modelled with the empirically based model rheology of
\cite{gdr2004}. The comparison of such computations with {our}
experiments and DEM simulations {will} probably act as a sensitive test
of the empirical rheology.

\begin{figure}
  \begin{center}
\includegraphics[width=\textwidth]{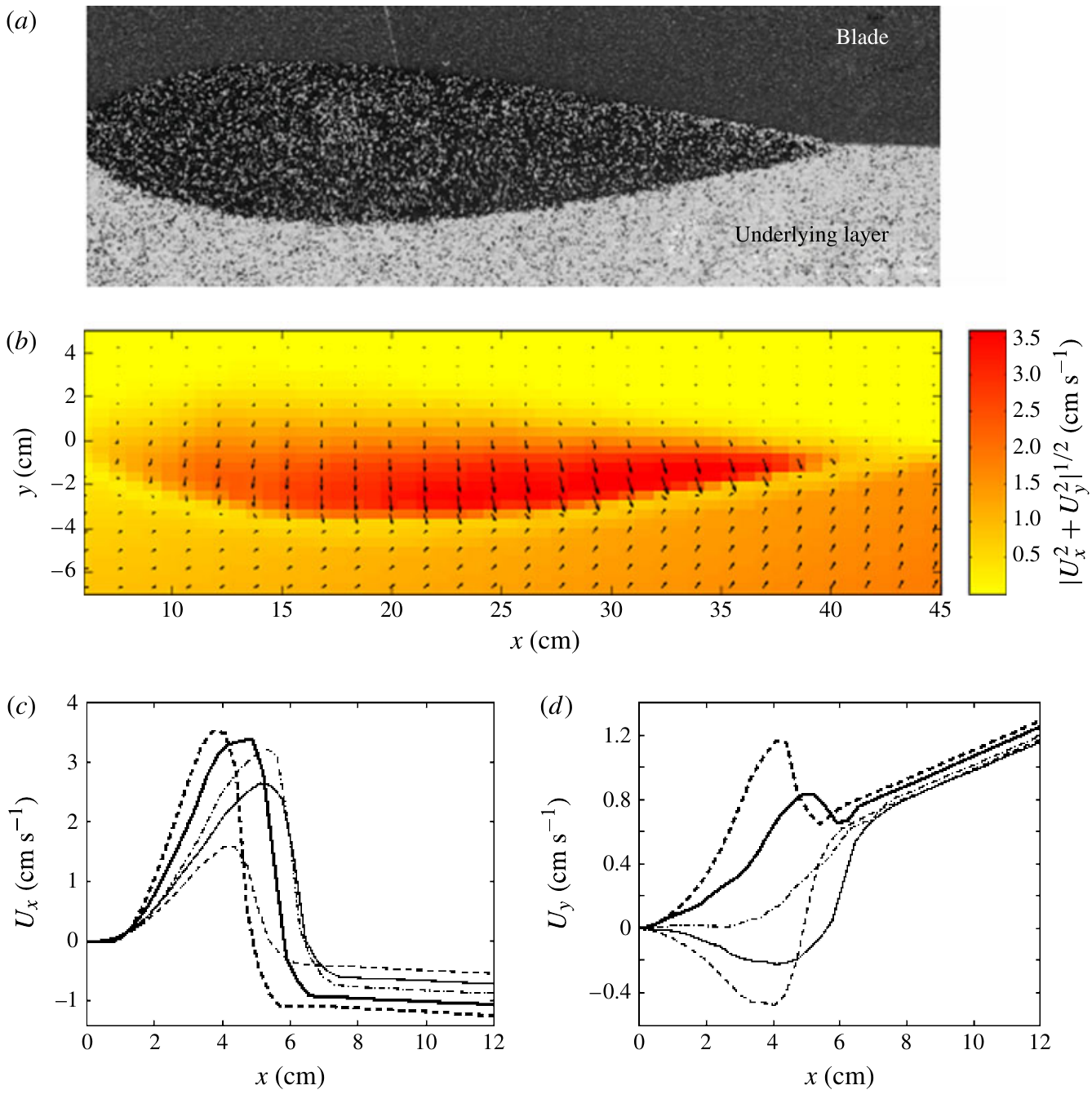}
    \caption{Surface velocity field of a dune of aquarium
      sand. (a) shows the raw image and (b) depicts the
surface speed in cm/s as a density over the plane of the image.
The surface velocity components perpendicular and parallel to the blade,
$U_x$ and $U_y$, are plotted as arrows in (b), and then in (c,d)
along the lateral positions $y=10$\,cm (dashed), $15$\,cm (continuous),
      $20$\,cm (dashed-dotted), $25$\,cm (thick continuous) and $30$\,cm
      (thick dashed).  Note that, in view of the inclination of the
      surface, the velocity field $(U_x,U_y)$ does not correspond to
      the horizontal Cartesian velocity $(u,v)$.  }
    \label{fig:profile_PIV}
  \end{center}
\end{figure}

From an experimental perspective, our approach has focused mainly on
recording the shapes of the free surfaces of bulldozed dunes.  Although
it is difficult to determine the full velocity field
within the dune non-invasively, it is certainly possible to measure the {surface
velocity} profile. Indeed, by using a mixture of grains of
different colours (in this case, black and white), we were able to take
some preliminary particle image velocimetry measurements for dunes of
aquarium sand, as shown in figure~\ref{fig:profile_PIV}.  The
avalanching flow at the front of the dune is clear in these
measurements, which provide a more demanding test of any theoretical
model.  The perpendicular surface velocity increases towards the front of
the dune, as the avalanching particles accelerate. There is also a
significant lateral surface velocity, which is probably due to the
overall surface shape.  The spatial structure of the flow
field warrants further investigation.

As a model granular flow, the two-dimensional bulldozer could be
considered as midway in complexity between the inclined-plane or shear-cell configuration and more complicated flows like the rotating drum.
The inclined-plane and shear-cell tests are steady uniform flows that
are free of any yield surfaces or rigid plugs. The streamlines in the
rotating drum, in contrast, piece together paths of acceleration
and deceleration, with flow bounded from below by a yield surface. The
wedges of our bulldozed flows do not appear to contain any yield
surfaces, but particles do experience
phases of acceleration and deceleration as they traverse the average
streamlines. Thus, the flow might well prove useful as a probe of
granular rheology, a stepping stone from the inclined plane or the shear
cell to the rotating drum.

\begin{figure}
  \begin{center}
    \subfigure[]{\includegraphics[width=\textwidth]{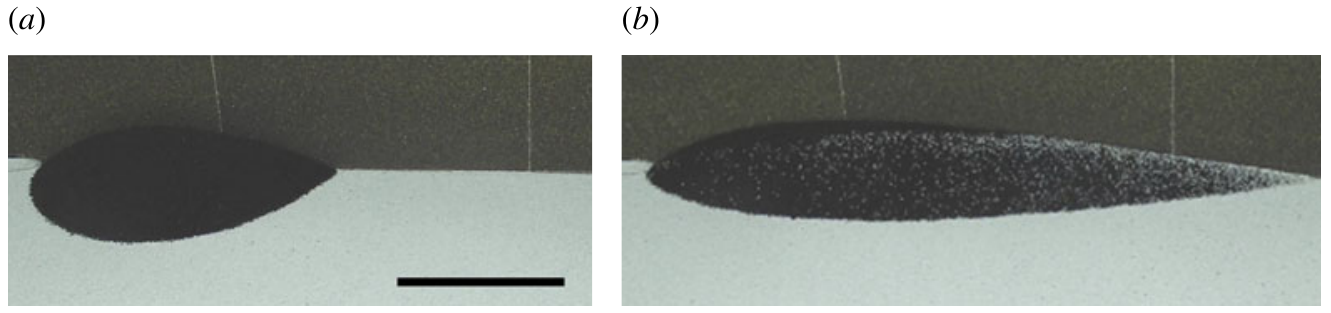}}
    \caption{Photographs from an experiment with $m=1000$ g,
      $\Omega=0.05$\,rad\,s$^{-1}$ and $y_0=25$\,cm and different
      coloured aquarium sand.  The initial pile of black sand was
      bulldozed over a layer of white sand.  Panel (a) shows the
      collision of the dune with the blade.  Relatively quickly, an
      exchange of particles between the sandpile and the underlying
      bed is visible, as shown in panel (b)  (at $t=35$\,s). Scale bar in (a) is $15$ cm.}
    \label{Num_Theo_profile_c}
  \end{center}
\end{figure}

Of course, although the mean flow of a steady two-dimensional
bulldozer possesses closed streamlines, the finite size of particles
and their granular temperature allows particles to cross transport
barriers and mix. Moreover, for our rotating bulldozer, the slower
lateral spreading also implies transport. Indeed, qualitative
experiments using aquarium sand of different colours {highlighted the
  exchange} of granular material between the initial dune and the
underlying layer; {as is apparent in} figure~\ref{Num_Theo_profile_c}.
Such mixing is also clear in the DEM simulations (figure
\ref{fig:dem3d}). Quantitative measurements and modelling of this
phenomenon are beyond the scope of the present paper but may be of
interest, for example, when on is considering the transport of pollutants
over soil and the avoidance of contamination is a critical objective.
A detailed exploration of particle migration would also be relevant to
granular segregation; the bulldozer being an interesting experimental
configuration in which to study this phenomenon.

Another interesting observation which we did not explore in detail was
the effect of slip over the surface of the rotating table.  When the
table was not covered by sandpaper, but was left as a smoother wooden
surface, the uniform layer initially emplaced clearly began to slide
during the approach to the bulldozer.  In particular, the flow divided
up into three characteristic regions. Well ahead of the dune, the
underlying layer remained flat and stationary, while closer to the
blade, the dune built up into a quasi-static avalanching mound that
appeared similar to those documented earlier. In between, the sliding
underlying layer created a buffer zone with a much shallower slope
than the dune behind it. Interestingly, the compression incurred over
this buffer appeared to generate surface features {reminiscent} of a
fine wrinkling pattern \cite[{\textit{cf.}}][]{melo}.

\section*{Acknowledgements} 
This work was initiated during the 2012 Geophysical Fluid Dynamics
summer program at Woods Hole Oceanographic Institution, which is
supported by the National Science Foundation and the Office of Naval
Research. We thank Anders Jensen for help in building the experimental
apparatus and Claudia Cenedese and Karl Helfrich for the use of their
equipment in the Coastal Research Laboratory. The DEM simulations were
performed using the Darwin Supercomputer of the University of
Cambridge High Performance Computing Service using Strategic Research
Infrastructure Funding from the Higher Education Funding Council for
England and funding from the Science and Technology Facilities
Council. We also acknowledge computing time on the CU-CSDMS
High-Performance Computing Cluster.

\bibliographystyle{jfm}

\bibliography{mybib}

\end{document}